\documentclass{aastex63}

\shorttitle{Big image classifications}
\shortauthors{Teimoorinia et al.}
\graphicspath{{./}{figures/}}

\begin{document}

\title{Assessment of astronomical images using combined machine learning models}

\correspondingauthor{Hossen Teimoorinia}
\email{hossen.teimoorinia@nrc-cnrc.gc.ca, hossteim@uvic.ca}

\author{Teimoorinia H.}
\affiliation{NRC Herzberg Astronomy and Astrophysics, 5071 West Saanich Road, Victoria, BC, V9E 2E7, Canada }
 \affiliation{Department of Physics and Astronomy, University of Victoria, Victoria, BC, V8P 5C2, Canada}

\author{Kavelaars, J. J.}
\affiliation{NRC Herzberg Astronomy and Astrophysics, 5071 West Saanich Road, Victoria, BC, V9E 2E7, Canada }
 \affiliation{Department of Physics and Astronomy, University of Victoria, Victoria, BC, V8P 5C2, Canada}

\author{Gwyn, S. D. J.}
\affiliation{NRC Herzberg Astronomy and Astrophysics, 5071 West Saanich Road, Victoria, BC, V9E 2E7, Canada }

\author{Durand, D.}
\affiliation{NRC Herzberg Astronomy and Astrophysics, 5071 West Saanich Road, Victoria, BC, V9E 2E7, Canada }
 
\author{Rolston, K.}
\affiliation{NRC Herzberg Astronomy and Astrophysics, 5071 West Saanich Road, Victoria, BC, V9E 2E7, Canada }
 \affiliation{Department of Physics and Astronomy, University of Victoria, Victoria, BC, V8P 5C2, Canada}

\author{Ouellette, A.}
\affiliation{NRC Herzberg Astronomy and Astrophysics, 5071 West Saanich Road, Victoria, BC, V9E 2E7, Canada }



\begin{abstract}

We present a two-component Machine Learning (ML) based approach for classifying astronomical images by data-quality via an examination of sources detected in the images and image pixel values from representative sources within those images.  
The first component, which uses a clustering algorithm,  creates a proper and small fraction of the image pixels to determine the quality of the observation.
The representative images (and associated tables) are $\sim800$ times smaller than the original images, significantly reducing the time required to train our algorithm. The useful information in the images is preserved, permitting them to be classified in different categories, but the required storage is reduced.  The second component, which is a deep neural network model, classifies the representative images. Using ground-based telescope imaging data, we demonstrate that the method can be used to separate ‘usable’ images from those that present some problems for scientific projects -- such as images that were taken in sub-optimal conditions.  This method uses two different data sets as input to a deep model and provides better performance than if we only used the images’ pixel information.   The method may be used in cases where large and complex data sets should be examined using deep models. 
Our automated classification approach achieves 97\% agreement when compared to classification generated via manual image inspection. We compare our method with traditional results and show that the method improves the results by about 10\%, and also presents more comprehensive outcomes. 

\end{abstract}



\section{Introduction}
\label{introduction}

Ground-based observational data necessarily varies in quality. Information regarding celestial objects is collected using a wide array of telescopes and instruments, often gathered and stored in a variety of different formats, such as astronomical images. The number of observations and variety of instrumentation create multiplicative complexity in astronomical data, including imaging. Different data sets are targeted for various science projects, and the impact and accuracy of the projects are directly linked to the quality of the data sets. On the other hand, a high-quality data set can be very complicated, with multi-dimensional characteristics. In general, complex problems need more complicated methods (or a set of well-arranged methods) to be solved. Images captured by a ground-based telescope, for example, are typically complex in terms of the range of objects they detect. The degree of complexity can also be highlighted when we take different observational conditions into account. 
Such conditions can, in turn, cover a broad range of different phenomena: from low-quality imaging caused by turbulent air and poor weather conditions, to images that suffer from transient detector-noise issues. 
The variations in image characteristics impact the utility of data.
Low-quality observations limit the confidence in science projects that use them — which can restrict reliance on such projects, making them less productive.

Classical methods and pipelines classify astronomical images based on  deterministic techniques. Different surveys such as SDSS, PanSTARRS and 2MASS have various ways of assessing image quality, based on the characteristics of the surveys. For example, quality assurance employed by 2MASS \citep{Skrutskie06} verifies the pipeline data-quality with an automated software system which can determine problems such as telescope-tracking and poor atmospheric conditions. The SDSS survey tracks a number of flags and quality parameters such as clear/cloudy, noise and PSFWIDTH. The latter is the Full Width Half Maximum  (FWHM) of the point spread function (PSF) in arcsec the median of which, in r band, is $1.3\pm0.2$ arcsec. A seeing of 2 arcsec is rare and may be a criterion for flagging an image. The problem of poor telescope tracking can also be registered by the initial processing of the data itself \citep{Ivezic04}. As another example, the Phase 2 Proposal Preparation (P2PP) Tool, which the European Southern Observatory uses, measures the minimum average seeing of an observation, and if the average PSF width is larger than a requested absolute value the observation may be repeated.

Classical and deterministic methods in surveys are usually cheap to run. For example, they can inspect some parameters such as ellipticity and FWHM and provide results that are easy to interpret. However, combining many different settings to make a decision based on the overall pattern in a survey needs more sophisticated methods \citep{teimoorinia17}.  Generally, classical methods are not able to combine the distribution of different parameters of a data-set to make a decision based on probabilistic models. In other words, in the conventional procedure, deciding (for example) whether an image is good or bad is independent of the quality of the rest of the images.

ML algorithms and methods (such as artificial neural networks, which can range from shallow to deep models ) have been shown to handle complex astronomical problems \citep{teimoorinia12,teimoorinia14,bilicki14,aghanim15,teimoorinia16,ellison16a,ucci18,teimoorinia18}.
They can predict different physical parameters and detect different patterns in different surveys \citep{rafieferantsoa18} or recognize various interesting astronomical sources, such as strong lensing candidates \citep{jacobs17,pourrahmani18,lanusse18}. Deep-learning methods are also used or exoplanet transit classification \citep{Ansdell18}  and  the non-parametric method Random Forest has been used to discover new nipper stars \citep{Hedges18}. 
ML approaches have also been used to measure the correlations between different galaxy structure parameters, revealing potentially serious problems in contemporary models of galaxy evolution  \citep{Bluck19}.  ML is finding an ever increasing role in astronomical data analysis.

There are various methods in ML; however, most break into two main approaches: ‘supervised’ and ‘unsupervised’. In a supervised method, we know what we want (the target, or ‘true’ answer), and our goal is a model that can achieve results as close as possible to the ‘true’ answer. In contrast, an unsupervised method has only the input data, and allows exploration of that data. 
One approach is to find clusters within the data by the construction of Self-Organizing Maps (SOM) \citep{Kohonen82, KO01}. In this method, the aim is to reduce the dimensionality of a data set and inspect the relationships between clusters of data points found in the set \citep[e.g., ][]{rahmani18}. Combining the two methods, supervised and unsupervised, can further increase the power of the ML approach. 

There are many cases in ML  where we can use our knowledge about a data set (feature-engineering processes) to make an algorithm work better. Other accepted methods (mostly, deep-learning ones) use less-engineered features as input. In both cases, however, a data set under study should be presented to a model in such a way as to make the selected model’s job easier. Generally, an astronomical image can be complex, and we cannot expect an ML model to learn directly from such complex data. Besides, a wide-field telescopic image, where a considerable fraction of the pixels contains empty sky, is not a good candidate as an input because the information density is quite low, although the sky by itself does contain some information of limited value. The process we present here extracts the most valuable image-quality classification features from our wide-field imaging data set. 

In ML one can also combine two fundamental approaches: supervised and unsupervised. Deep clustering is an approach that combines techniques and has seen significant development in computer vision applications \citep{Caron18}. For example, \cite{Xie16} use an unsupervised system  which itself consists of two unsupervised methods. This method (which is called deep embedded clustering) is a combination of a K-means algorithm and a deep Autoencoder, in which the latter can consist of different CNN layers  \citep[for deep clustering methods also see][]{Al18}. In this work, we implement a model that combines an unsupervised technique with a supervised deep model to solve a complex problem in astronomical imaging\protect\footnote{The code is available in:  \protect\url{https://drive.google.com/open?id=1vKrWgjiYKqDXlfs_RnOnkjxeMh1dcS9D}}: automated quality assurance. 



An important topic in machine learning research is to improve the performance of the models used.  The major contribution of this work is to present a novel and creative dimensionality reduction step in which image pixel data can be fed into a deep learning model in an effective way. In this way, we show that combining suitable models and data sets can significantly improve the overall performance of the methods used individually, both in accuracy and speed.

To demonstrate our approach, we have selected an image-quality assessment problem in which the ultimate goal is to evaluate images and separate high- and low-quality images (i.e., those that are usable for science projects from those that are not). The applicability of the method, however, is not restricted to assessing image quality. Here, we deal not with a binary problem (acceptable/unacceptable) but a multi-class classification problem with different categories of `bad' (low quality) images.  We will show that a combination of different ML methods allows an automated solution to this complex problem.

High-quality astronomical images enable researchers to achieve more impactful outcomes. For example, Megacam images are of limited use individually and combining multiple exposures, on the same field, increases the sensitivity and reduce image defects.  These `stacked' images provide more suitable data for broad classes of research.  
Our method enables assessment of the individual Megacam images so that the highest quality inputs can be selected for stacking.  The Canadian Astronomy Data Centre (CADC) uses the output of the combined model presented in this paper as the input to MegaPipe \citep{SG08}.

In Sec.\ref{sec:data} we describe the data chosen in this work and in Sec. \ref{sec:method} we explain the method and combine two ML methods. Sec. \ref{sec:result} presents some results of our image classification. In Sec. \ref{sec:discussion} we present a short discussion, and we compare our results with those using classical methods in Sec. \ref{sec:comparison}. The summary will be presented in Sec.\ref{sec:conclusion}.

\section{Data}
\label{sec:data}

\begin{figure*}
\centering
\includegraphics[width=14.cm,height=14cm,angle=90]{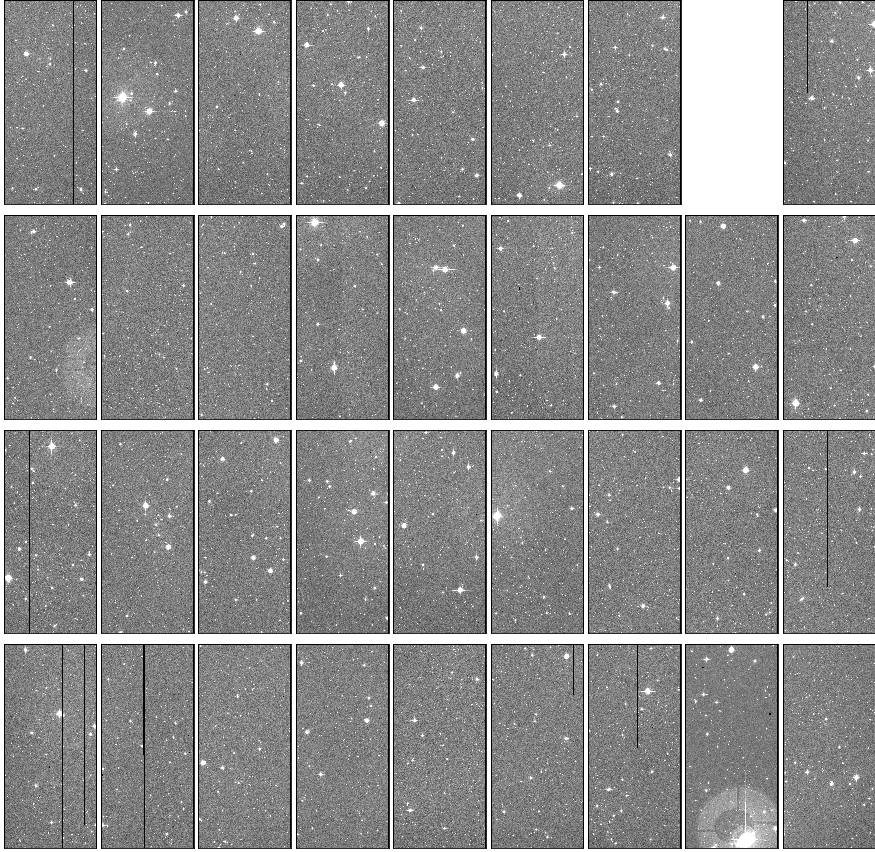}\\
\vspace{.5cm}
\includegraphics[width=7.cm,height=14cm,angle=90]{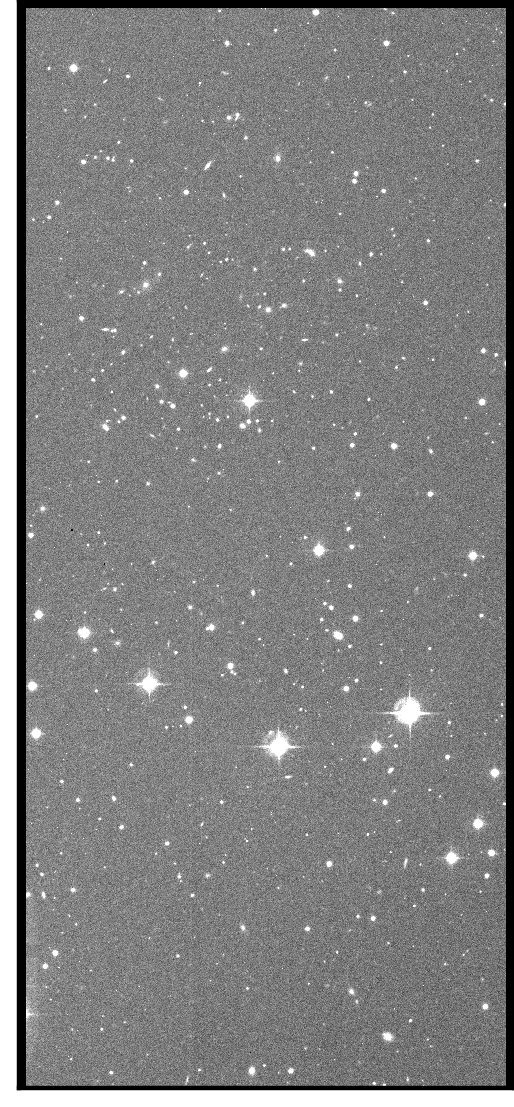}
\caption{The top plot shows an example  of a typical MegaCam exposure consisting of 36 CCDs. The white section of the image represents a non-functioning CCD caused by sporadic electronic failures. The bottom plot shows one of the 36 CCDs (the upper right one;  $4644\times2112$ pixels) in a larger view.}
\label{fig:36good}
\end{figure*}

For this analysis, we explore our assessment method using selected images from the MegaCam instrument mounted on the Canada-France-Hawaii Telescope (CFHT) \citep{Boulade03}. 
MegaCam is an optical imaging mosaic CCD camera consisting of 40 ($2048\times4612$ pixels)  CCDs/images, with each pixel imaging $\sim 0.184$~arc-second region of sky. 
An individual exposure is stored in a single Multi-Extension FITS file of approximately 800~MBytes size.
During initial operation, each MegaCam image consisted of 36 CCDs four of which are vignetted and thus not recorded.
Recently, however, a new set of camera filters has expanded the unvignetted field of the camera, and all 40 CCDs are recorded for each image. 
An example 36-CCD exposure is shown in Figure~ \ref{fig:36good}. 
Optical images taken by ground-based telescopes, such as MegaCam, exhibit a wide range of image quality. 
Low-quality images contain problems regarding telescope-tracking (from slight to severe), poor sky conditions (e.g., poor visibility; cloudy conditions) and different background issues, such as pronounced background fluctuations, severe object-saturation, and non-astrophysical background patterns. 
Our goal is to develop an ML-based method that automatically ranks or groups images based on quality characteristics.

The MegaPipe processing system \citep{gwyn08} at CADC  calibrates and combines CFHT MegaCam images that are first passed through a visual inspection to classify them into different image-quality based categories. 
The visual classification categories are ‘good’, really bad tracking (RBT), bad tracking (BT), various problems in the background (BGP), and poor visibility during observation (B-Seeing). 
Besides, during MegaCam observations there were times when not all the CCDs in the mosaic were functioning, these exposures have a secondary quality value of ‘Dead-CCD’ (see Figure~\ref{fig:36good}). 
Over 5,000 MegaCam exposures (i.e., $\sim 5000\times36$ CCDs) have been visually assessed and placed into one of the categories above. This initial quality assessment allows us to use human experience to build models that will enable our ML-based process to place images into the classes mentioned above.  

\begin{figure*}
\centering

\includegraphics[width=7.cm,height=7cm,angle=0]{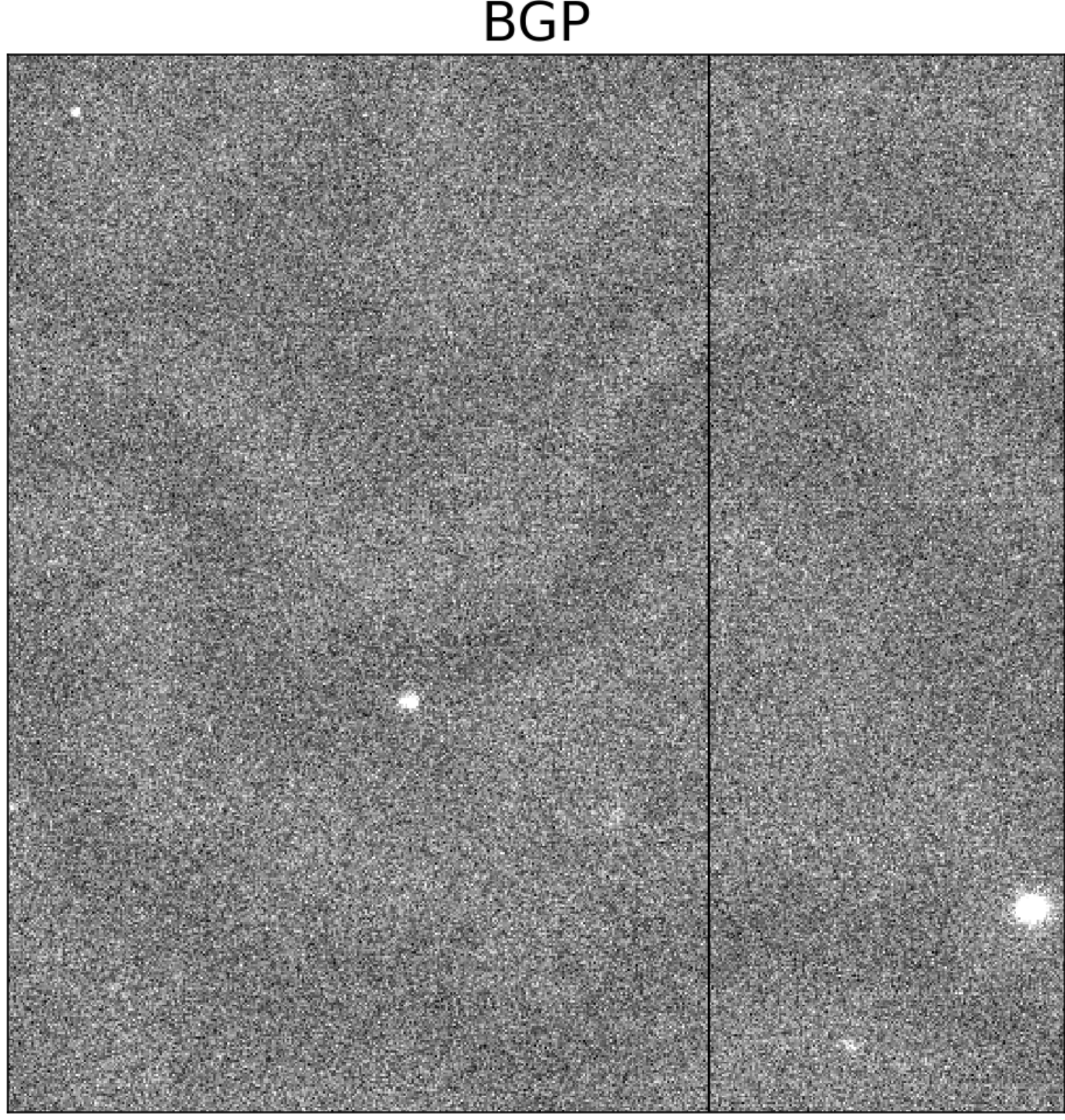}
\includegraphics[width=7.cm,height=7cm,angle=0]{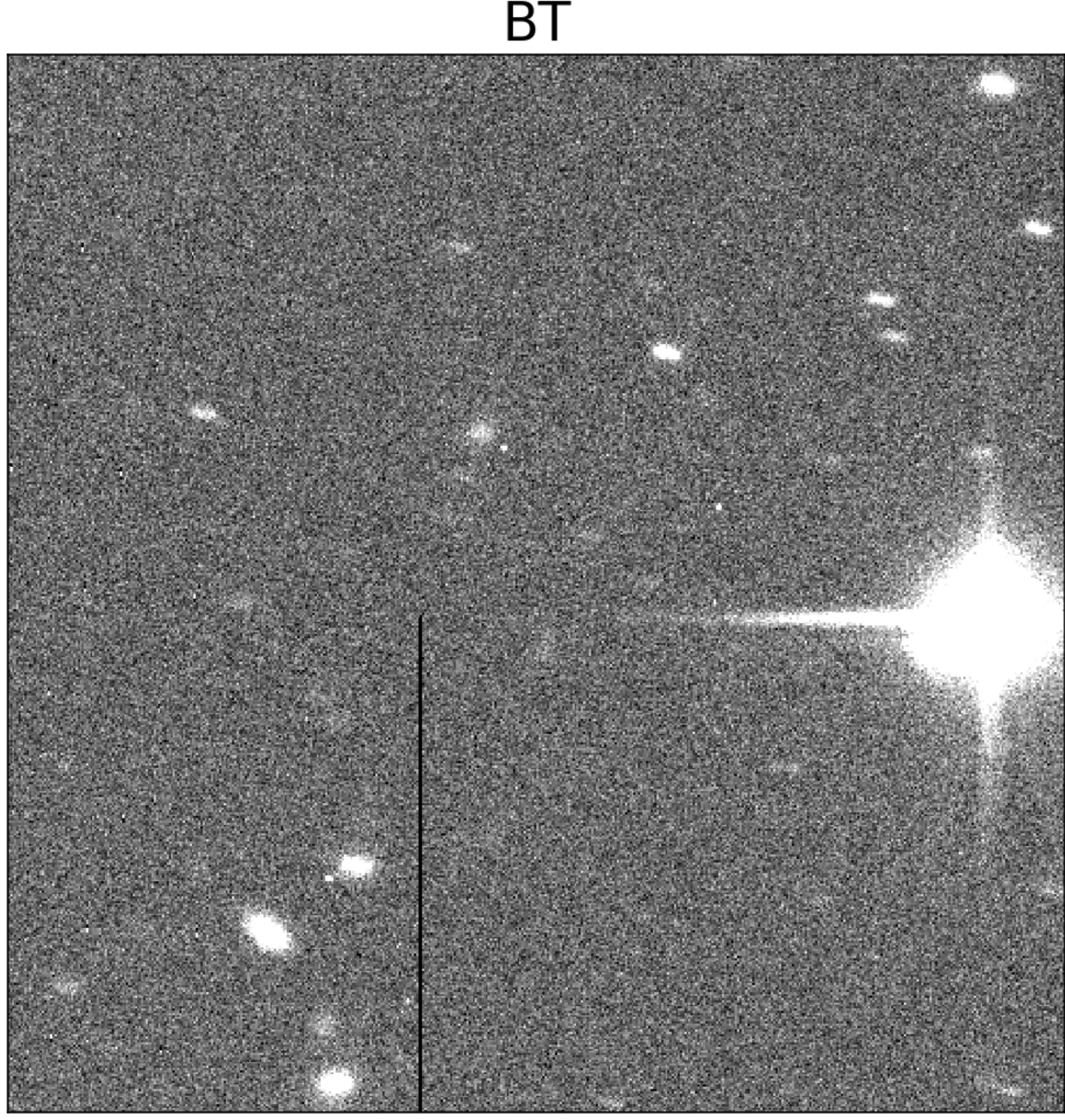}
\includegraphics[width=7.cm,height=7cm,angle=0]{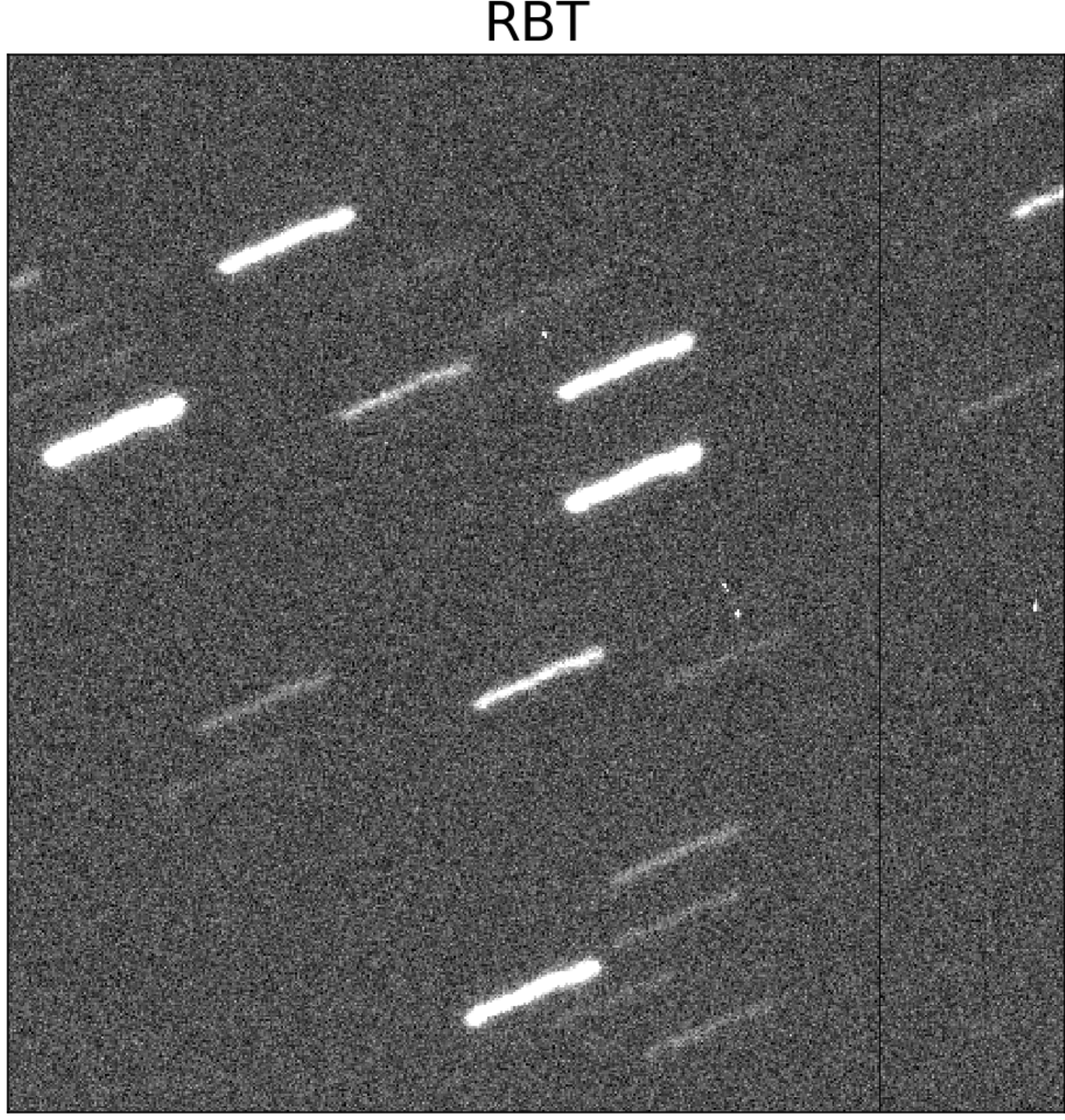}
\includegraphics[width=7.cm,height=7cm,angle=0]{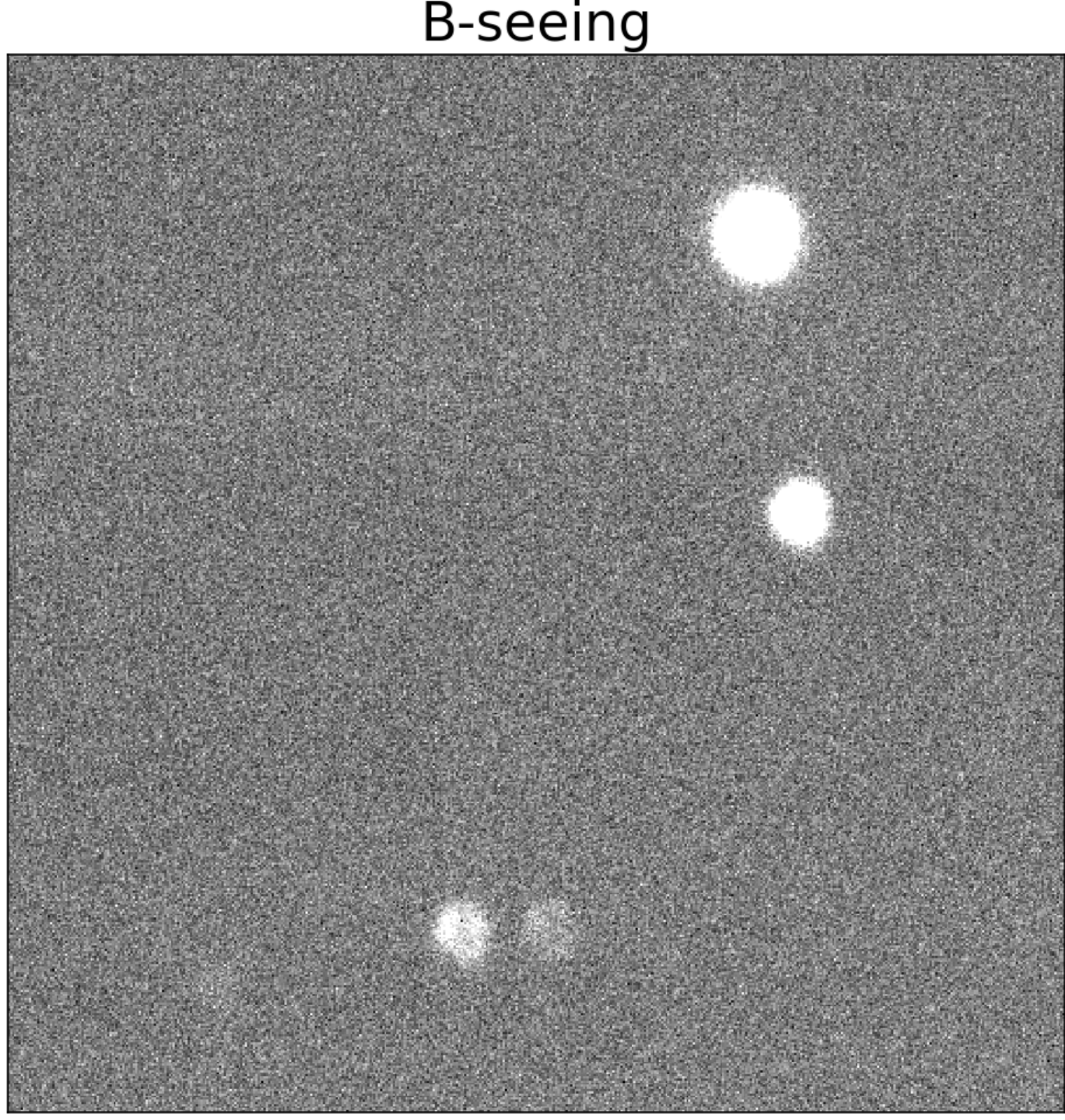}
\includegraphics[width=14.cm,height=7cm,angle=0]{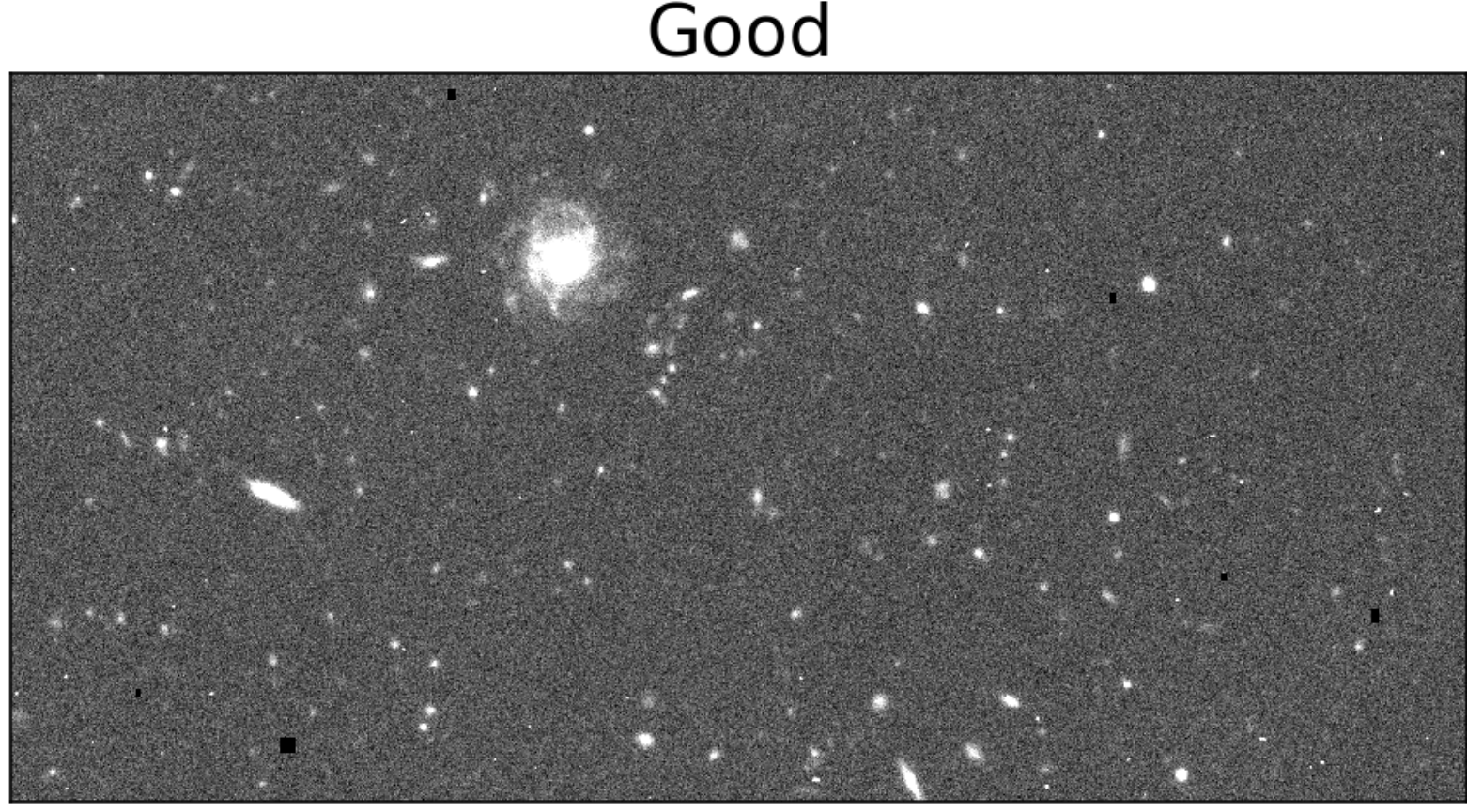}
\caption{Five different targets for our models in this paper. They include images with different problems in the background (BGP), bad tracking (BT), really bad tracking (RBT), bad seeing or bad observational conditions (B-Seeing) and an instance of a Good image at the bottom of the figure.}
\label{fig:target}
\end{figure*}

To generate a suitable training set, we randomly selected some exposures from all the available classes and re-examined all the images to make sure we had correct labels in our training set. 
Some examples of all the classes are presented in Figure~\ref{fig:target}. To see the detail of exposure we show only a small (magnified) part of a single CCD. 
As can be seen from Figure~\ref{fig:36good}, each exposure contains a large quantity of information.
One of our goals is to introduce a method where only a small fraction of the complete image is directly input into our ML algorithm. 

\section{The Method}
\label{sec:method}

In ML approaches, data should be presented in an appropriate way for the model. 
The presentation method depends on the nature of the problem under study. 
As mentioned in Sec.~\ref{sec:data}, CCD pixel values (from an exposure) can require significant memory resources to examine. 
Feeding many 100s of such large exposures into a deep-learning model would be computationally expensive and may not be the optimal analysis approach.  As an example, one way to avoid expensive computations could be to cut  20 small images randomly out from a CCD (as representative images) and then feed them into a network. In this way, there is no guarantee of having different sources with different characteristics in the small cut-out images. In a visual inspection, we usually search for various objects in an image to assess its quality. In other words, the random method can ignore (or over-highlight) the real character and nature of the image. Here, we present a method to extract a small, representative example of an image in such a way that the useful information from the larger, more complex image will generally be preserved. A clustering method can help to render a fair representation of an image.

\subsection{The representative images}
\label{sec:representative}

First, we use SOURCE EXTRACTOR (SE) \citep{bertin96} to detect astronomical sources within the images and determine parametric quantities using the pixels associated with each source recognized. In our SE analysis, the following parameters are returned: ISO0 (Isophotal area at level 0), ELLIPTICITY (1 - B\_IMAGE/A\_IMAGE),  BACKGROUND (background at centroid position),  and CLASS\_STAR (star/galaxy classifier output). We also record the image header keyword EXPTIME, as an additional input to train our model. SE measurement catalogues were created for a randomly selected set of training exposures chosen from each of the classes defined in section~\ref{sec:data}. Each of the individual SE catalogues was then concatenated to form a global source catalogue. Using $\sim3,000,000$ randomly selected sources from SE produced the global source catalogue we train a Self-Organizing Map (SOM; SOM; which is a kind of artificial neural network). The SOM of SE output parameter measurements form the initial step of our classification process.

\begin{figure}
\centering
\includegraphics[width=10.5cm,height=10cm,angle=0]{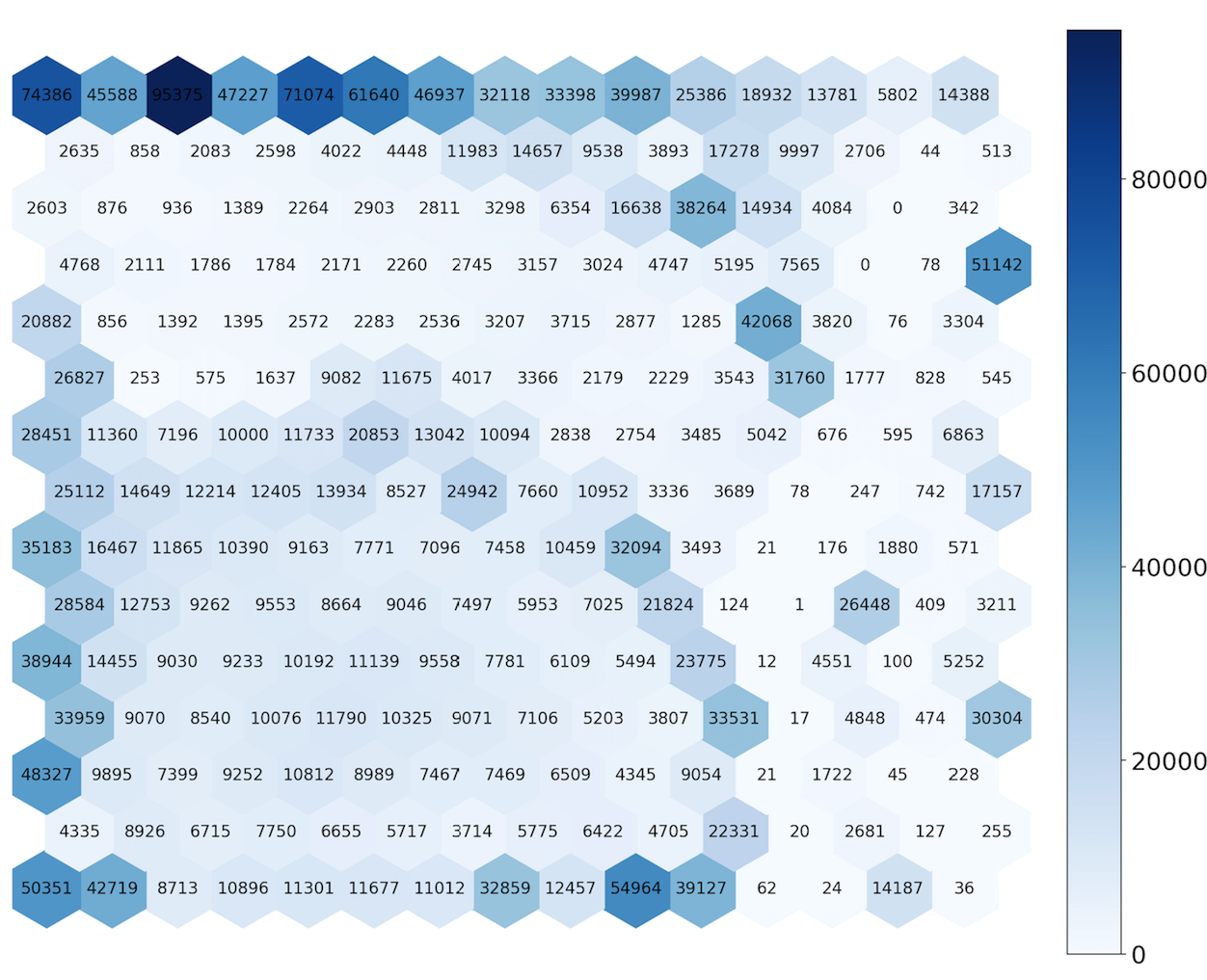}
\includegraphics[width=10.5cm,height=11cm,angle=0]{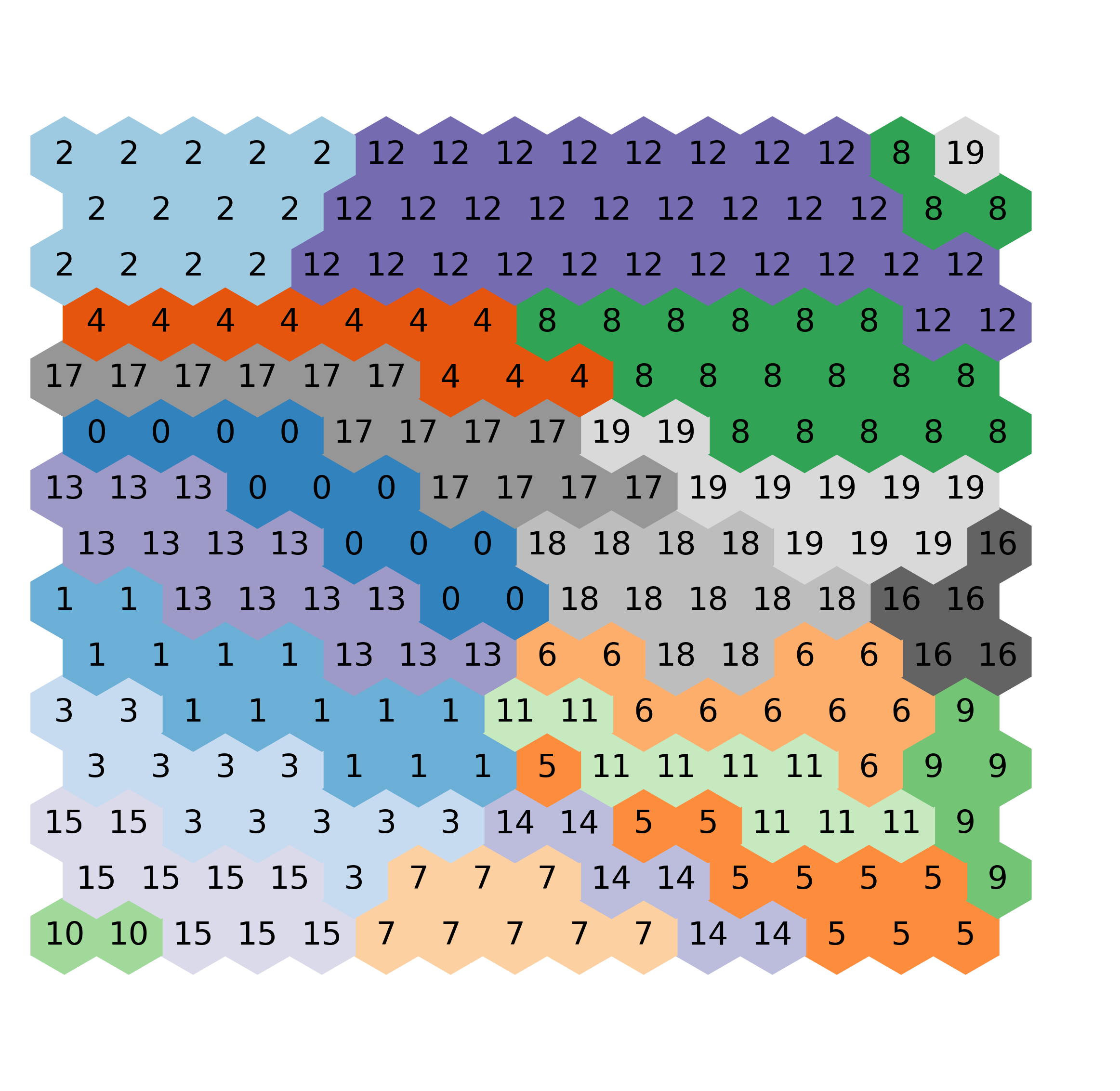}

\caption{The top plot shows a hit map.  Here, 3,000,000 sources, with five parameters, are used to train the map (by the SOM algorithm) and then the sources can be arranged and distributed among the 225 nodes. In this way we train a SOM network. For example, cluster 10 contains 50351+42719 sources. After the training step, the 225 nodes are clustered into 20 clusters (by the K-means method of SOMPY) and given labels from 0 to 19. Each new input image source list is then distributed into these clusters.}
\label{fig:cluster20}
\end{figure}

A SOM places each source into a two-dimensional ($\rm{N} \times \rm{N}$) map based on similarities or dissimilarities to other sources in the map while attempting to preserve the topological properties of the input space by using a suitable neighbourhood function. The two-dimensional map contains a grid of $\rm{N}_u$ ($= \rm{N} \times \rm{N})$  units (or nodes) in which each unit is represented by a d-dimensional vector unit, $\rm{\vec U_i}$.  The units on a map can have rectangular or hexagonal forms in which the latter can create more connections with adjacent units (e.g., the top plot of Figure \ref{fig:cluster20}; $\rm{N}_u= 15\times15$). First, vector units are randomly initialized. Then, an input datum $\rm{\vec X}$ (which is also a vector with dimension d; here d=5) is compared (e.g., by a Euclidean function) to all the units on the map. The algorithm is looking for the best matching unit (BMU) for the input, which is obtained by finding the minimum value of $\left \| \rm{\vec U_{i}- \vec X} \right \|$, (i=1 to $\rm{N}_u$).  So, an input vector can find its BMU on a map with $\rm{N}_u$ units. In the first iteration (t=1), this method is repeated for all input vectors, X$_{k}$ (here; k=1 to 3,000,000).  So, all 3,000,000 sources can be ‘distributed’ (not necessarily uniformly) on a map of size $\rm{N} \times \rm{N}$. 

The next step is to update the vector units. This step can be done by:

\begin{equation}
\rm{U_{i}(t+1) = U_{i}(t) +\alpha~f_{bi}\times (X-U_{i})}\\
\label{eq:update}
\end{equation}

In Eq.\ref{eq:update}$,~ \alpha$ is the learning rate  and f$_{bi}$ is the neighbourhood function that is concentrated on the BMU and gives the distance between the BMU and neighbouring units. A Gaussian neighbourhood function is defined by:

\begin{equation}
f_{bi} = e^{{{ - \left \| {r_b - r_i } \right \|^2 } \mathord{\left/ {\vphantom {{ - \left( {x - \mu } \right)^2 } {2\sigma ^2 }}} \right. \kern-\nulldelimiterspace} {2\sigma ^2 }}}
\label{eq:gaussian}
\end{equation}

In Eq.\ref{eq:gaussian},~$\sigma$ is the neighborhood radius that can be chosen by the user (usually as a fraction of N, here $\sigma=3$). With an error function \citep[which can be defined as a function of multiplication of Euclidean distance and the neighbouring function e.g.,][]{KO91} and after several iterations the sources are distributed on the (trained) map.  In this work, we have chosen N=15.  So, after training the network, we have a ($15\times15$) map in which each unit contains similar objects with specific characteristics. When the training step is finished,  the number of similar instances in each unit can be counted. The top plot of Figure \ref{fig:cluster20} shows the distribution of the 3,000,000 sources on the map (i.e., the hit map).  Such a map generally does not have specific axis labels (because it is created by dimensionality reduction methods). For example, in this instance of the SOM, the node at the top right of the grid contains 14,388 sources with similar characteristics based on the four selected SE parameters plus the exposure time of the image.

Different, but connected, nodes in the SOM contain sources with more or less similar characteristics. The goal of the SOM network is to cluster sources together based on similarities in the provided features. When the size of a map is large, related units need to be grouped. \citep{Vesanto00}. To implement the SOM algorithm, we use SOMPY \citep{sompy18} in which clustering is done in two separate steps. First, the input data is distributed (on a 2D map) by SOM (i.e., the dimensionality reduction round). Next, (i.e., after SOM is trained) the SOM is clustered. Different methods can be considered for implementing the second step, such as hierarchical agglomerative clustering and partitive clustering using K-means \citep{Vesanto00}. SOMPY uses a K-means algorithm in which the SOM can be clustered into $\rm{n}_c$ different clusters. In a map 15x15, for example, $\rm{n}_c$  can be selected a number between 1 and 255. SOMPY also uses  a batch version of the algorithm where the learning rate is not used \citep[e.g.,][]{KO95}. In this work, each iteration takes $\sim13$ seconds to finish (with a 2.8 GHz Intel Core i7 computer). After 200 iterations no improvement in the error function is seen, and the model is ready to use. The response of the trained model to new data is rapid.

It should be mentioned that the K-means clustering algorithm could be applied directly on the SE output to cluster sources into $\rm{n}_c$  categories. However, it does not take any topological information to arrange the data. SOM is a more complicated and robust method than K-means. SOM, in the limit, can turn to the K-means algorithm \citep[e.g.,][]{rahmani18}. After we arrange the 3,000,000 sources on a map by SOM then K-means can be an effective method to group the nodes on the map.

The SOM clusters can be used to group sources into classifications. One cluster might contain point sources while another contains extended sources, for example. Between those clusters, there can be intermediate cases. In this way, we can create a classification space into which any new source can be mapped (based on the values of the measured parameters). Within this space, each image will have a different density distribution. An image of mostly star-shaped objects, for example, will have a high density of sources in a different area of the SOM compared to an image containing mostly extended objects. In a  $15\times15$ map  with 3,000,000 instances, we have more than (on average) 13,000 sources per unit, which is a sizable number considering the five parameters taken from SE. In other words, we train our network with a large random sample of all available data; space is considered to be ‘complete’; and we can map data from a new image onto this space.

Using the trained map, we take catalogues containing the same parameters measured for sources in a new image (obtained by SE) and distribute those sources onto the trained network. In this way, even without seeing the image (with examining the hist map and properties of the clusters, see Sec. \ref{sec:clusters}),  we can find the image characteristics (rich in stars or perhaps containing many galaxy clusters, for example) based on the distribution of sources on the map. In other words, there are patterns in the maps which we can use to recognize the image content. Although this is not the explicit goal of the current investigation, the patterns in maps can then be used to search for images with particular kinds of content, without reference to specific catalogues of such sources or knowledge of the sky-location contained in the image. The map places sources from a given image into groups with similar source-parameter values, and those sources are then categorized based on which cluster they appear in.

We have chosen to have 20 clusters within our SOM (i.e.,  $\rm{n}_c=20$), with a map of size $15\times15$. In the bottom plot of Figure \ref{fig:cluster20} we show the same map that is grouped into 20 clusters.   Each detected source of the new image is placed into a particular cluster. The shape of the distribution (how many sources of the cluster into which groups) depends on the nature of the image under study. In the next step, we randomly select a source from each of the 20 SOM clusters as a representative source of that cluster for the given image. We then extract pixel values for those representative sources (image subsets); we refer to this as a ‘cut-out’ of the source. In this way, we have 20 cut-outs for each input image. Using the hit map, we also can count the number of sources a particular image has in each cluster of the SOM. Using the 20 cut-out image subsets we create a ‘representative’ image, and the number of sources from the image contained in a given SOM cluster is the weight assigned to the particular image cut-out. The SOM provides a method for mapping the content of the image and allowing significant compression without losing information concerning the global quality of the image data.

The size of the map and the number of clusters in a map are free parameters. We have used a map $15\times15$ showing 20 clusters (i.e., Figure~ \ref{fig:cluster20}). In this way, by choosing a $15\times15$ map (i.e., a higher resolution than a $5\times5$, for example), we find a more complex structured map; the 3,000,000 sources have more phase-space to be arranged on the map. In all cases, however, the same sources were used to train the map with the same five parameters as input. Selection of the map resolution and the number of clusters is done through an iterative process to achieve the desired classification, as determined by the end goal of the process. In this case, finding a model that accurately predicts the ‘quality’ of the image was the goal that harmonized the free parameters of map size and the number of clusters. As was mentioned (in a map with 15x15 nodes) $\rm{n}_c$ a number between 1 and 255 can be selected. The bigger the number of clusters, the better representative of an image will be however, more memory, a longer image processing time, and stronger computational sources would be needed.

\begin{figure}
\centering

\includegraphics[width=18cm,height=11cm,angle=0]{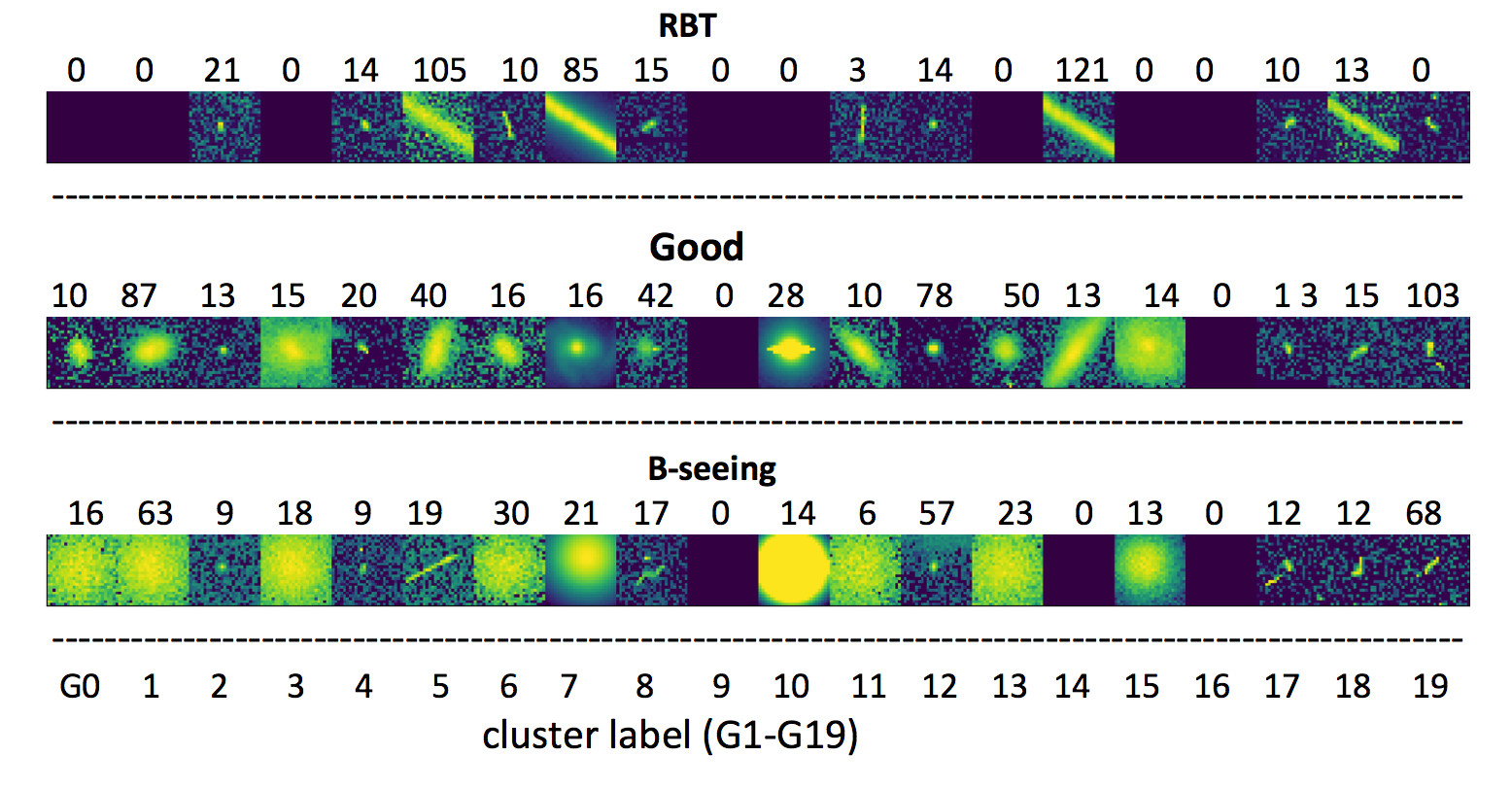}
\caption{Three different representative images of three different CCDs from different classes. Each representative image contains 20 cut-out sources from the 20 different clusters shown in Figure~ \ref{fig:cluster20} (from 0 to 19). The number of similar sources, in the corresponding cluster, can be seen on top of each cut-out source. For example, for the ‘good’ image in the figure, the cluster with label 19 contains 103 similar sources. }
\label{fig:three_ccds}
\end{figure}

Figure~~\ref{fig:three_ccds} shows representative sources from each of three different images in the different classes (from the training set). Each representative image contains $25\times25$ pixel cut-outs (postage stamps) from the 20 different clusters shown in Figure~~\ref{fig:cluster20} (numbered from 0 to 19). On the top of each postage stamp (here; $25\times25$ pixels) the number of similar sources in the corresponding cluster can be seen. For example, for the ‘good’ image in the middle panel, the cluster with label 19 contains 103 similar sources. So, for each image, we have created a representative image, with size $20\times(25\times25)$ pixels,  as the main input for a deep-learning model. We use the set of numbers in the hit map of each image as auxiliary data to train our image characterization model.

\subsection{The properties of the clusters}
\label{sec:clusters}
\begin{figure}
\centering
\includegraphics[width=8cm,height=4cm,angle=0]{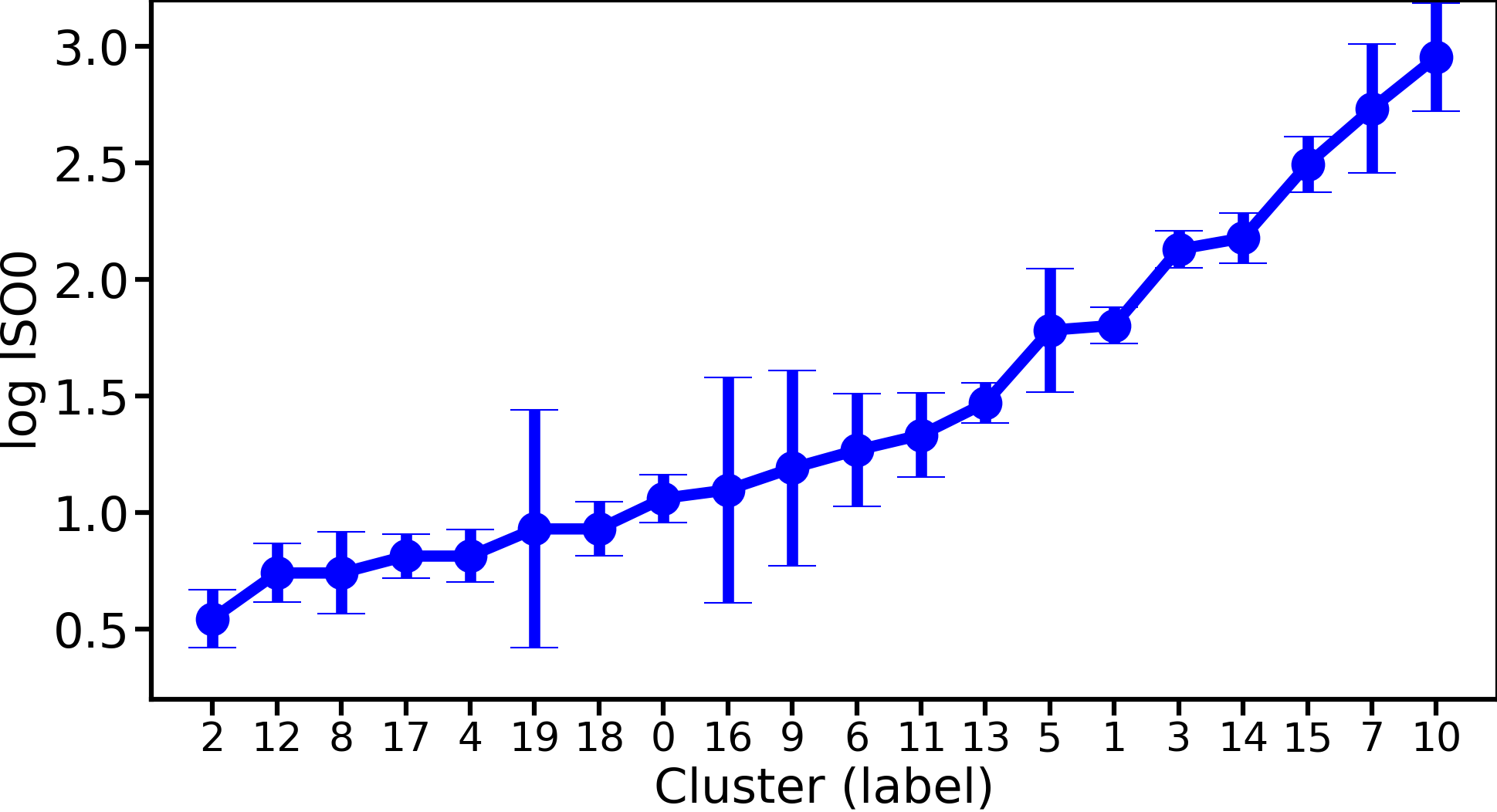}
\includegraphics[width=8cm,height=4cm,angle=0]{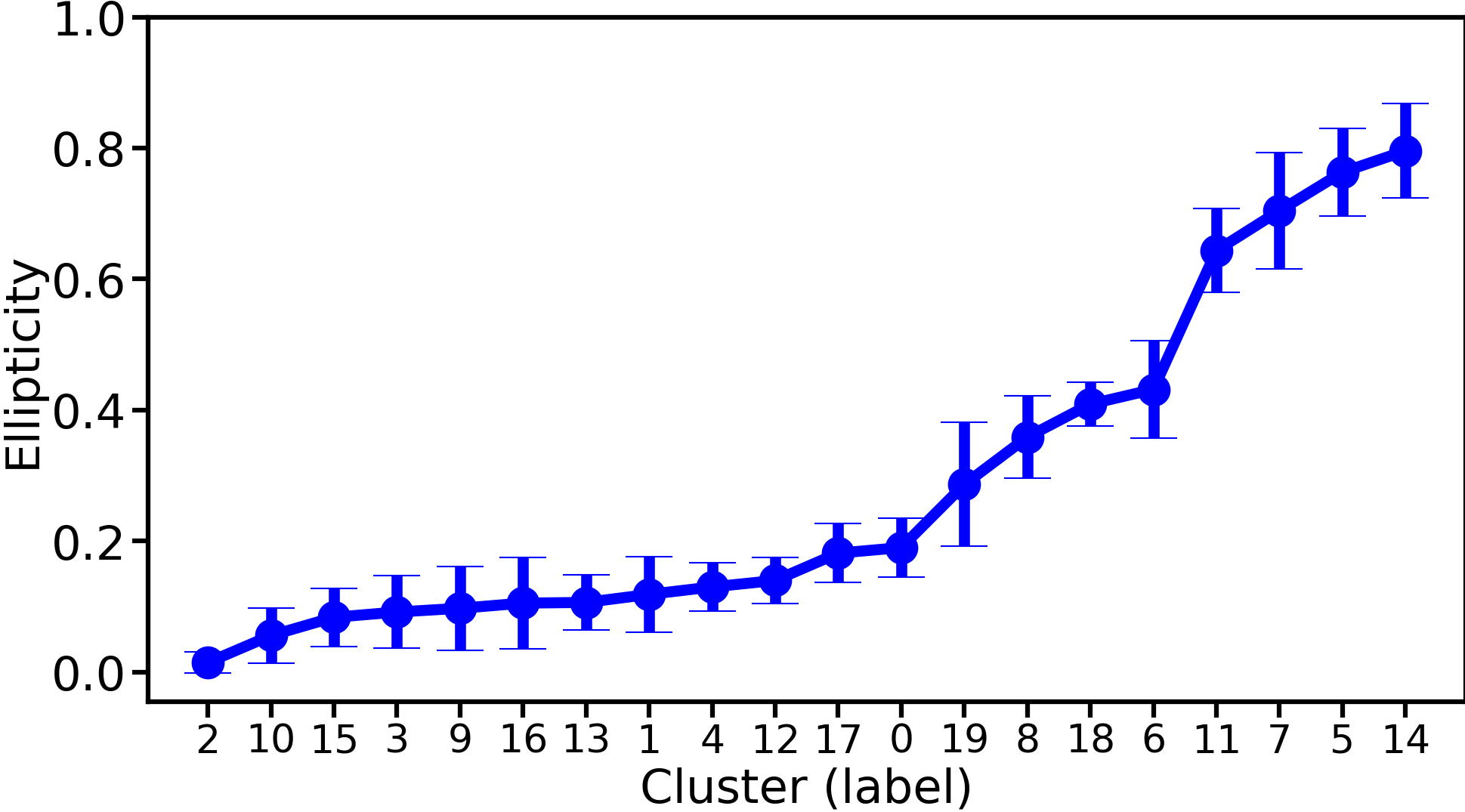}
\includegraphics[width=8cm,height=4cm,angle=0]{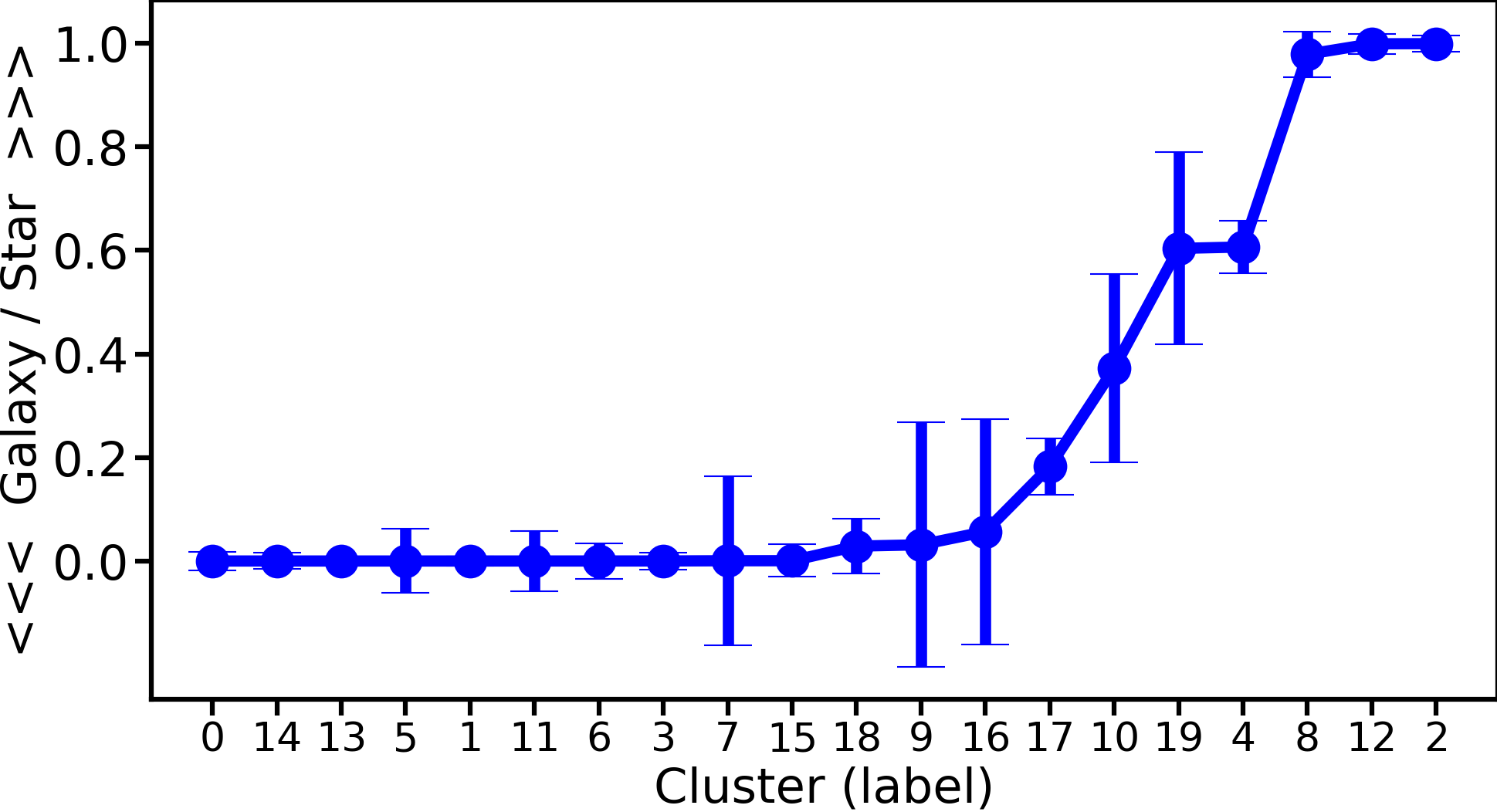}
\includegraphics[width=8cm,height=4cm,angle=0]{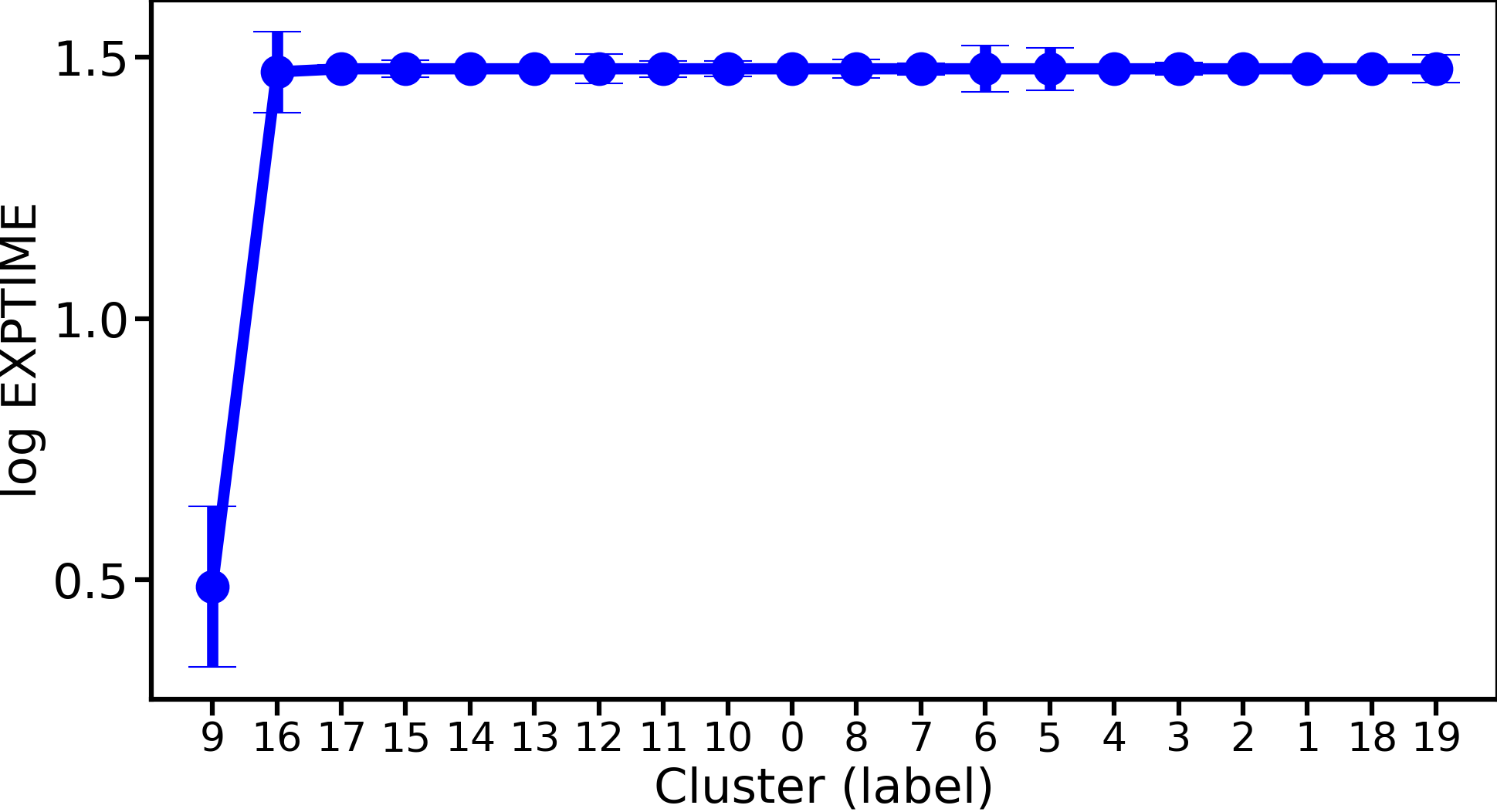}
\includegraphics[width=8cm,height=4cm,angle=0]{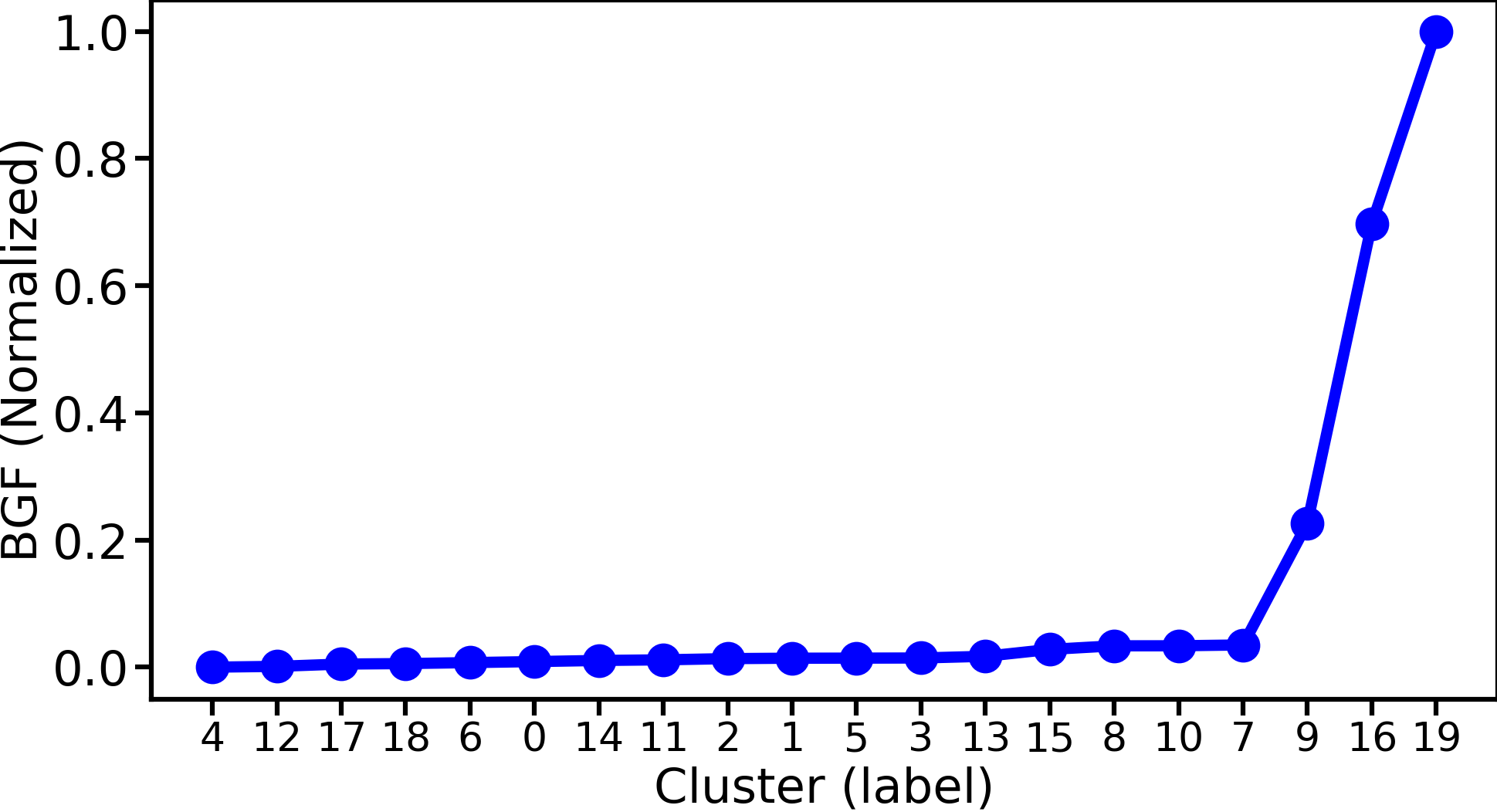}
\caption{  The  plot show the average value of the 5 parameters in each cluster.  For example, In the right top plot, cluster 2 contains round sources, and cluster 14 contains the sources with the highest elongation, on average. The four other plots are related to the parameters ISO0,  star/galaxy, the parameter regarding EXPTIME (exposure times greater than 30 are set to 30 to highlight low exposure cases), and background fluctuation around the mean value. For example, cluster 9 has the lowest exposure time and cluster 19 has the highest background fluctuation within the clusters.}
\label{fig:weights}
\end{figure}

In this section, we present quantitative plots of the average value of the SE parameters in each cluster (see Figure \ref{fig:weights}). As mentioned in previous sections, each cluster in the bottom plot of Figure 3 contains similar sources. From the top left plot of Figure \ref{fig:weights}, we can see that, for example, clusters 2 and 10 contain respectively the smallest and largest sources in terms of their ISO0 values. In the plot related to EXPTIME, we have set all values of EXPTIME above 30 equal to 30 (see Sec. 5 for more details). In this way, we highlight low EXPTIME sources. For example, we can see that cluster 9 contain sources with low exposure time and also with relatively high background fluctuation (BGF). Again, cluster 2 contains sources with small ISO area, low ellipticity, EXPTIME$>=$ 30, and low BGF (which are detected as ‘stars’ by SE).

\begin{figure*}
\centering
\includegraphics[width=13cm,height=22cm,angle=0]{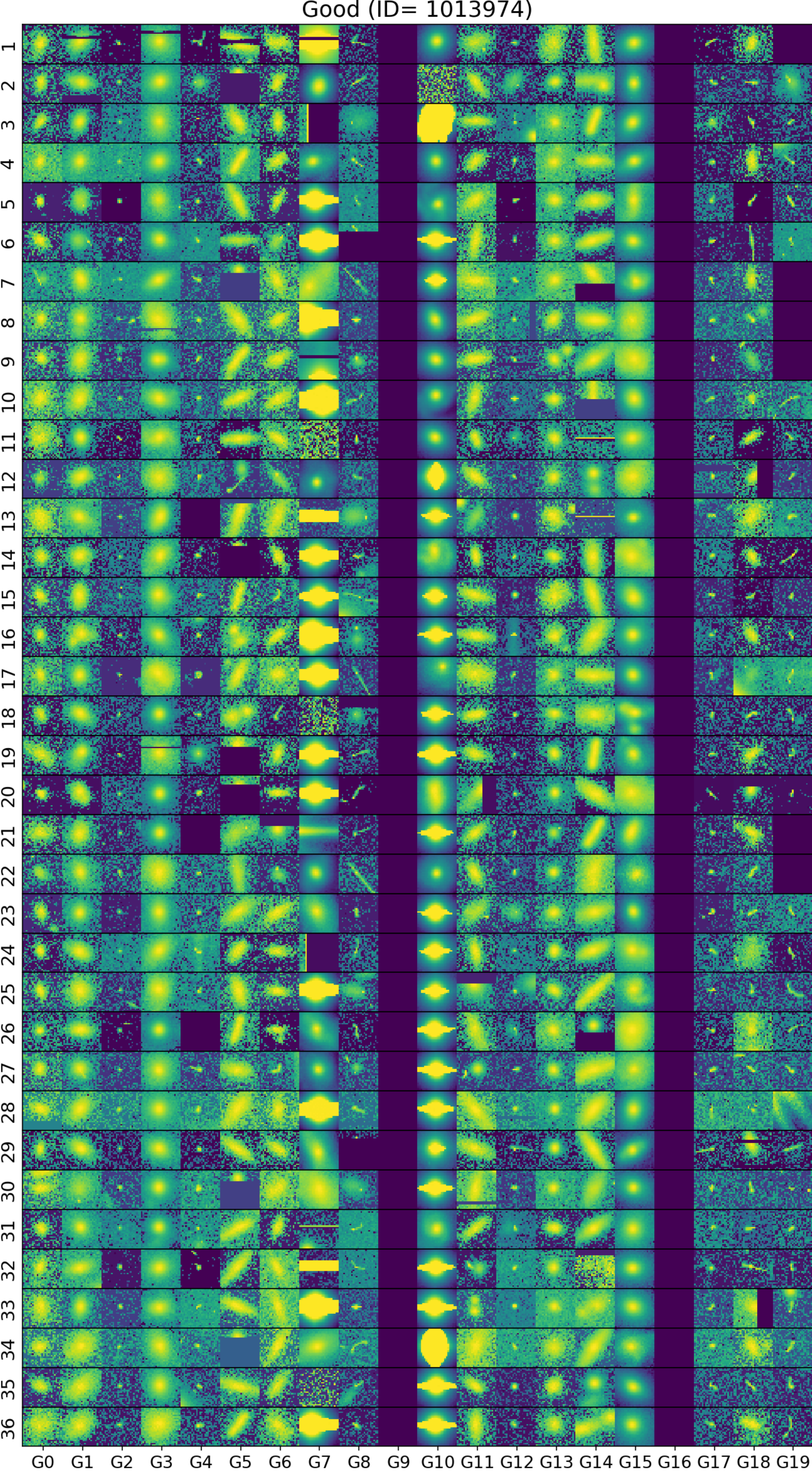}
\caption{The plot shows 36 representative images (CCDs) from one exposure (on top of each other), each with 20 cut-out images (25x25 pixels) taken from the 20 clusters shown in Figure~ \ref{fig:cluster20} (with labels 0-19). The plot is related to a 'good' exposure ($\sim800$ times smaller than the exposure). For example, cluster 3 is completely populated, which contains relatively large sources with low ellipticity, and relatively a high value of ISO0 (see also Figure \ref{fig:weights}).  However, for a RBT image (see Figure \ref{fig:36_2}) the cluster should be  empty, As another example, there is no source for the images in cluster 9 in Figures \ref{fig:36_1} and \ref{fig:36_2}, because both had high exposure times.}
\label{fig:36_1}
\end{figure*}

\begin{figure*}
\centering
\includegraphics[width=13cm,height=22cm,angle=0]{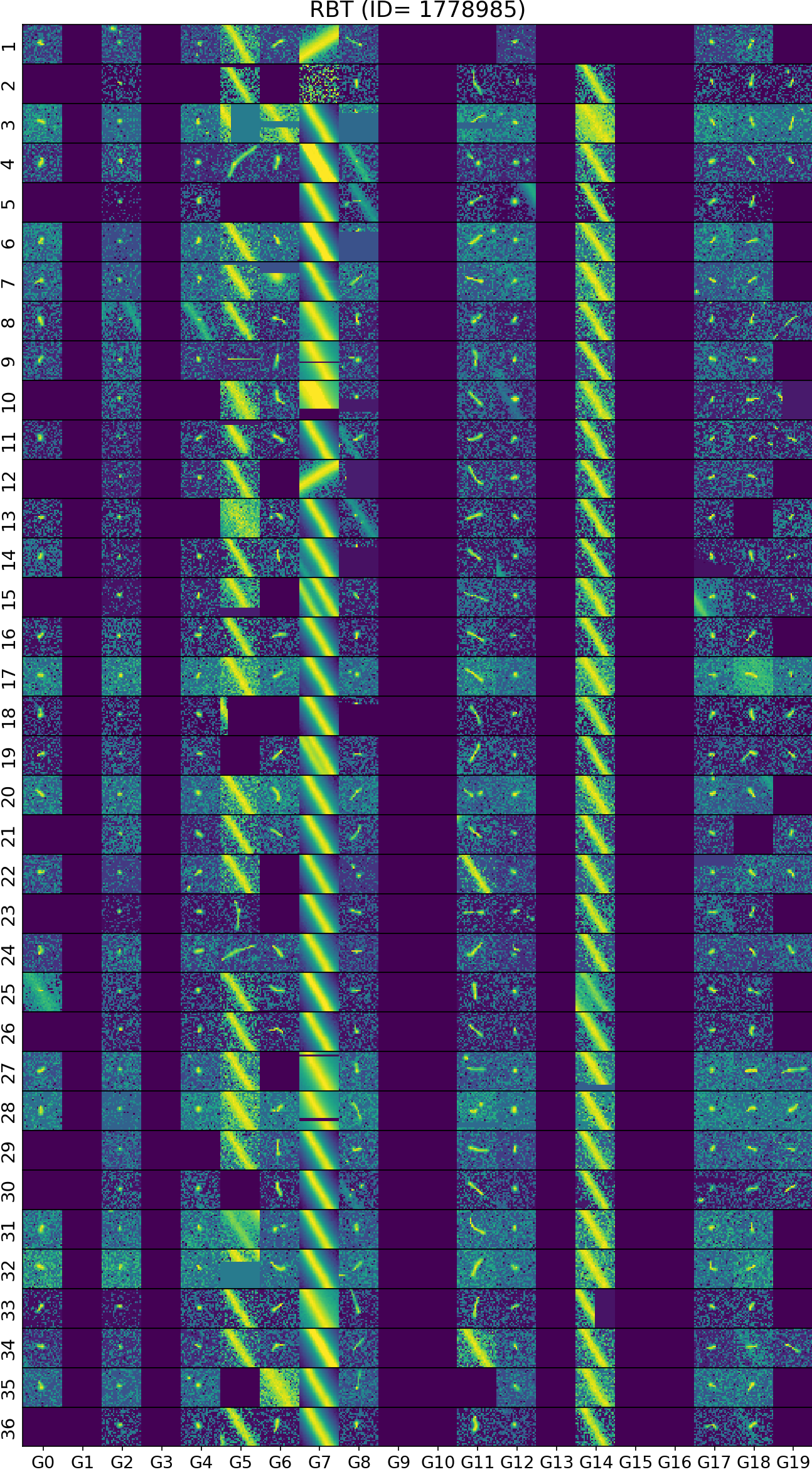}

\caption{The same as Figure \ref{fig:36_1} for a RBT image.}
\label{fig:36_2}
\end{figure*}

By examining the representative cut-out images for each SOM cluster for given images we find patterns that relate to image quality. For example, in Figure~\ref{fig:36_1} we present a representative image cut-outs for each SOM cluster (columns) for each of the 36 camera CCDs (rows) for a 'good' different exposures.  In Figure~\ref{fig:36_2} we  presents an RBT exposure. Different patterns can be seen. For example, cluster 9 is empty in both exposures because they are high-exposure instantiates, and cluster 9 contains sources measured in exposures with low exposure times. Another example, cluster 3 (which contains low ellipticity sources and relativily high ISO vlaues) in the RBT exposure is nearly empty (RBT exposures generally have high ellipticity). While for the good exposure, cluster 3 is well populated. These representative images are $\sim800$ times smaller than the main exposures as an input to a model, allowing rapid model training on low-memory machines. Besides this pixel information of different exposures, we also have statistical information describing the relative population statistics for each cluster in the $15\times15$ SOM. Plots such those shown in Figures~\ref{fig:36_1} and \ref{fig:36_2} (along with the statistical information) act as a ‘fingerprint’, and such patterns can be found in different exposures taken in different conditions which allow the quality-assessment to occur. We will train a (combined) deep network to explore these patterns and predict the quality of the image. We will then use these sets as a combined input to a combined model.

\subsection{The combined model}

Here we show that a combination of different ML methods can improve total performance while speeding up the learning and validation steps. We combine a SOM with a deep-learning model to classify images into five quality groups. We find that a SOM provides a method to select representative data as suitable inputs into a deep-learning model. Our goal is to provide a 'big picture' view of the data rather than explain fine-grained and technical aspects.
The intent of this manuscript is not to present a detailed code that can be broadly distributed; instead, the goal is to show that combining different ML models as well as available and relevant information can dramatically improve the performance and accuracy for the problem under study. The way of combining the information may depend on the nature of the problem under study. Here, we present an example of image classifications in astronomy. Deep-learning methods are being widely used in different areas \citep{Goodfellow-et-al-2016}, and the use of a SOM to organize the selection of classes of input is likely generalizable to those problems.  

There are a variety of different deep learning models publicly available. We use Keras \footnote{https://keras.io}, which is a high-level neural networks API and allows users to build complex CNN models quickly. In the previous section we have described the unsupervised part of our model (SOM),  here we briefly describe the CNN part.

A CNN model  can consist of different layers and take two-dimensional image input. In each layer, an image is convolved with a set of two-dimensional filters (usually the size of few pixels, e.g., 3x3).  The weights associated with the filters are randomly initialized.  The user can select the number of filters in each layer.  So, for $\rm{N}{_f}$ selected filters in the first layer, we will have  $\rm{N}{_f}$ convolved images.  These images are called feature maps.  A pooling function can downsize the feature maps in the first layer.  This step reduces the size of the feature maps by taking the maximum value of four adjacent pixels.  So a 2x2 max-pooling function, for example,  can reduce the number of pixels in a feature map by a factor of 4.  This procedure is essential to avoid using too much system memory.  The (reduced) feature maps made in the first layer will become input images for the second layer. In this way,  useful information from the image can be extracted using multiple network layers.  In this work, we use five CNN layers (see model M1 in the top plot of Fig. \ref{fig:perform}).  The output of the last layer will be fed into a fully connected layer.  However, before this step, all the two-dimensional image pixels in the last layer should be converted into a one-dimensional (flattened) array. Only flattened  data can be fed into a fully connected model.  Then the model can classify the data. In a fully connected layer,  each neuron in one layer is connected to all neurons in the next. In a neural network system, a neuron receives input from M other units (i.e., from the previous layer). The output of a simple neuron (e.g.,  in a layer of the fully-connected part) is:

\begin{equation}
\rm{output_i= F(b_i + \sum_{j=1}^{M} w_{ij} \times input_j)}
\label{eq:ann}
\end{equation}

In Eq.\ref{eq:ann}, w$_{ij}$ and b$_i$ are trainable parameters  and F is an activation function. The function, which has a non-linear characteristic, helps to capture the nonlinearity of the problem under study. In the CNN and the fully connected layers in this work, we use the activation function ‘relu’ \citep[a rectified linear activation function; e.g.,][]{Nwankpa18}. However, the last layer of the fully connected layer should be the softmax activation function. This function can convert an n-dimensional vector of the final layer (here, n = 5; i.e., the number of the classes) to a probability distribution so that the summation of the probabilities is 1. In other words, the output of the fully connected part is passed to the softmax function to produce the five classification probabilities. Between the layers, to avoid  over-fitting problems  one can also use different regularization function. We use a weight-decay regularization \citep[i.e., a L2 regularization; see chapter 3 of][]{bishop:2006}.   A more detailed discussion about deep learning and SOM methods can be found in the references presented in Sec.~\ref{introduction}.

\begin{figure*}
\centering
\includegraphics[width=18cm,height=9cm,angle=0]{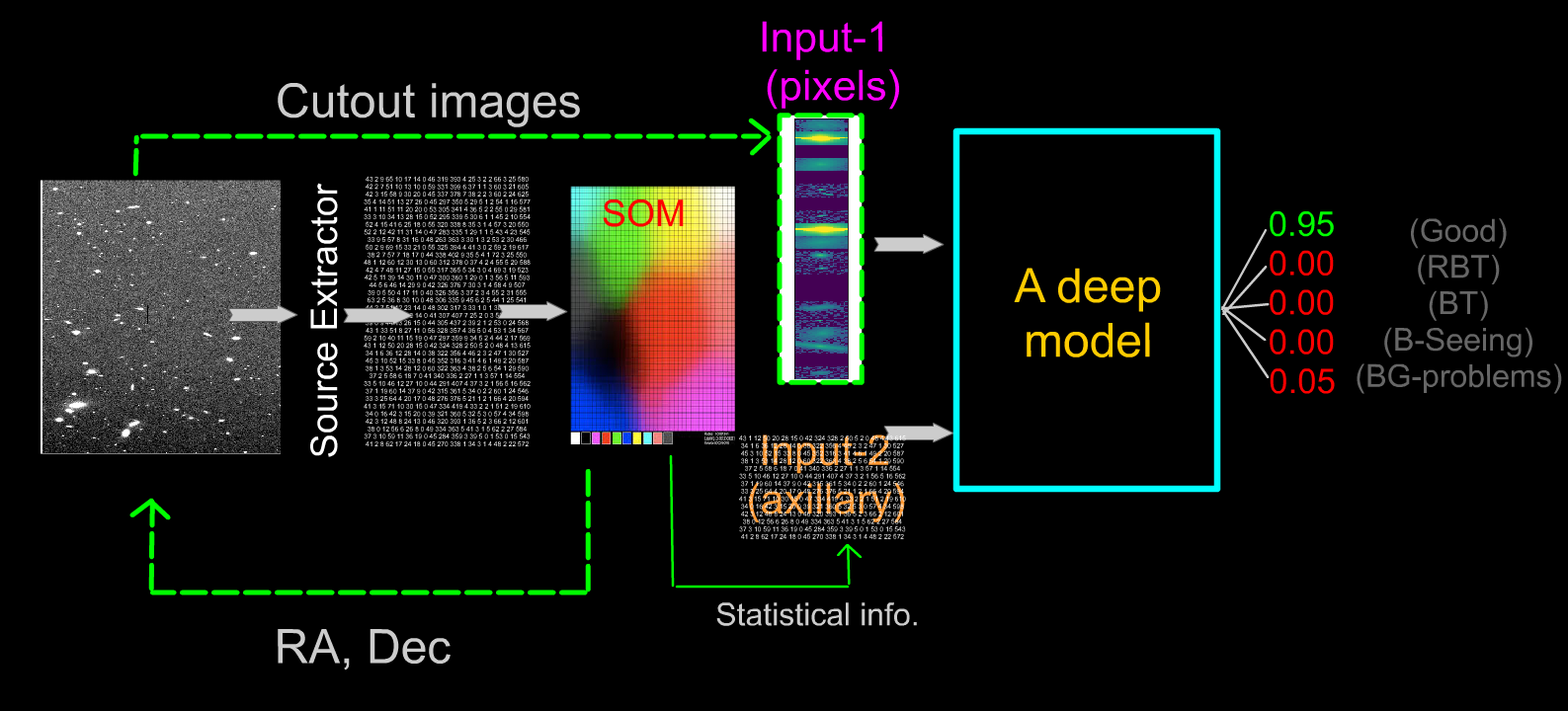}
\vspace{.5cm}
\caption{The plot is the combined model used in this paper: the selected parameter of detected sources from an image (the left image) can be extracted by SE, and the table from SE is then fed to a trained SOM. The SOM model can cluster the table into 20 clusters, and then we can pick up one object from each cluster. After that, a representative image (i.e., 20 cut-out objects using RA and Dec provided by SOM), using the main image, can be constructed. The representative image (i.e., pixel information) is the primary input (Input-1). Besides, we can obtain statistical information from SOM (the number of similar objects in different clusters). That is more information we will provide to the deep model (Input-2). The five classes are the output of the last model.}
\label{fig:combined}
\end{figure*}

Figure~\ref{fig:combined} presents a schematic view of our ML mode. First, we use the SE programme to extract useful information concerning the ‘sources’ detected in the image (the left side of Figure~\ref{fig:combined}), this provides parametric measurements for the sources. This information is fed to a SOM that has been trained using a big sample of the detected sources drawn from a selection of images of different quality classes: here, 3,000,000 sources. The SOM model places the sources into N cluster. The number of clusters is an option of the analysis, we found N=20 provided an effective end result. We then select one representative source from each cluster in the SOM. Using the original image pixels associated with each of these sources we construct a {\rm representative} image (i.e., an image built by connecting together the 20 sub-images, each centred on the representative source for that cluster). The representative image (pixel information) is denoted as Input-1 in the Figure. We also have available statistical information from SOM (the number of similar objects in different clusters). This statistical information is additional information that we provide to the deep model (Input-2). The target of the deep model is the five different probabilities described in Sec. \ref{sec:data}. Different models can be explored in the deep learning part of the process. Here we examine three different deep models: M1, accepts Input-1 only, M2 takes Input-2 only, and M3 which accepts both Input-1 and Input-2.

\section{The Results}
\label{sec:result}
In this section, we use the three models discussed in the last section (i.e., M1, M2, and M3), and compare the associated performances. Some examples of the predictions of the image classification are also presented.

\begin{figure}
\centering
\includegraphics[width=9.85cm,height=6cm,angle=0]{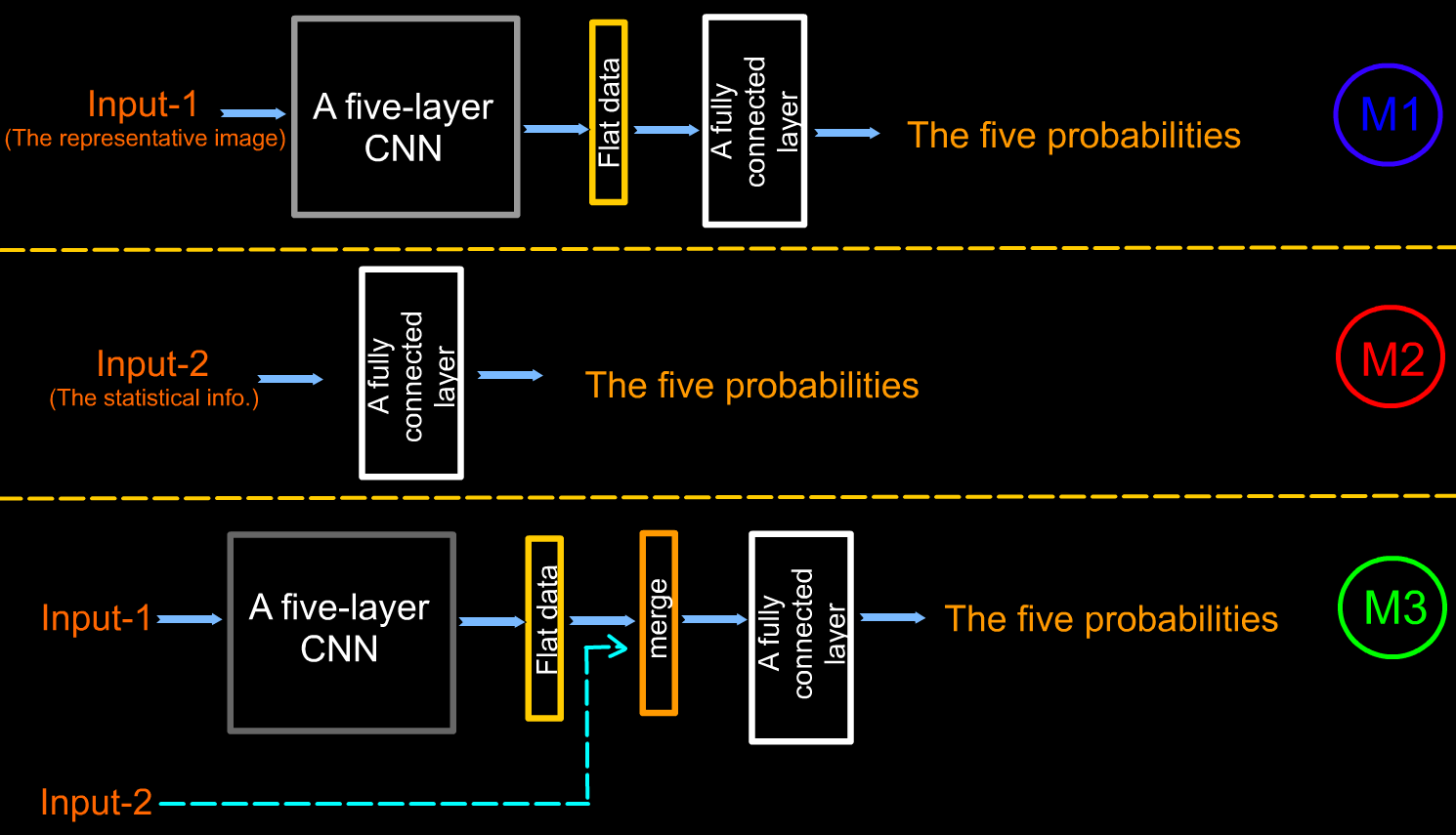}
\includegraphics[width=10.2cm,height=6cm,angle=0]{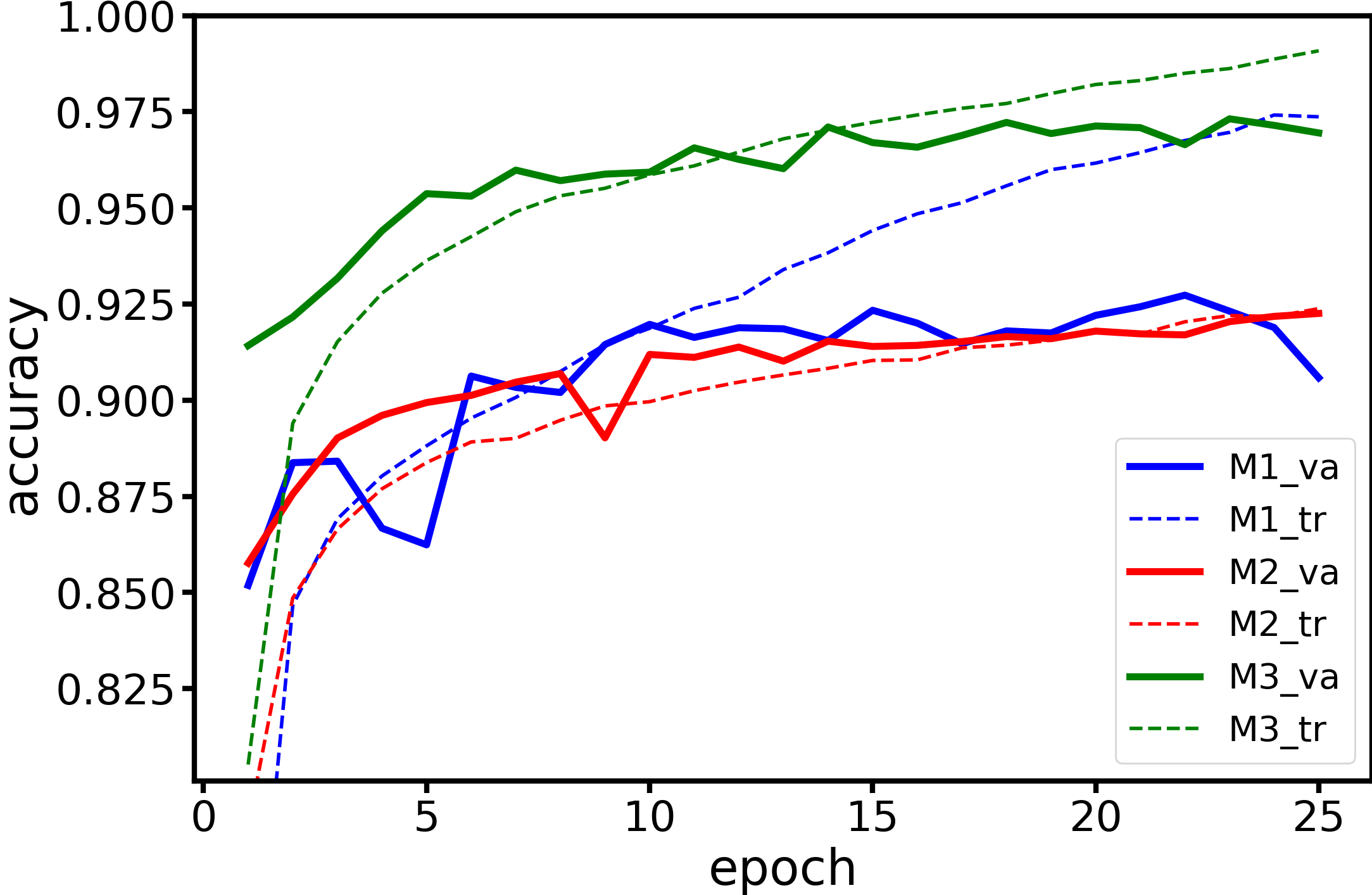}
\caption{The top plot shows the detailed picture for the deep models, i.e., M1, M2, and M3. We can use three different inputs for the models. M1 indicates a case in which only the representative image (Input-1) is fed to the model. M2 shows a model in which the input is just statistical information from SOM (Input-2). M3 shows a combination of the two inputs. The performance of the three models is shown for both training and validation sets. The performance of M3 shows a significant improvement over the other two models which use only one input.}
\label{fig:perform}
\end{figure}

In the top plot of Figure~\ref{fig:perform} we show the three different models that can be used, separately, to classify the images. In the top model (that denoted by M1), the representative image (Input-1) is provided to a five-layer CNN. Then, after converting to flat data, we use a fully connected layer to obtain the five probabilities. In the middle, M2, a deep model (only the fully connected layer) driven off Input-2, is shown. M2 gives the probabilities for each class directly from the distributions of the measured SE parameters (i.e., the statistical information from SOM). Using M3, we combine the output of CNN with Input-2 (which can be represented as flat data) and the merged data set is fed to the fully connected layer to obtain the probabilities. The three different models allow us to examine the influence or importance of the two types of content (parameter distribution and pixel values) for the end result. To select a suitable model (of all the three models presented in Figure \ref{fig:perform}), we train different models with different powers (more/fewer layers and more/fewer neurons in each layer) to make sure that we obtain a maximum accuracy/performance for the validation set. Adding more complexity/layers to model M2, for example, does not change the accuracy and can present over-fitting.

To examine the efficacy of our models we use separate training and validation data sets. We provide the sources, lists and sub-images for 60,000 pre-classified images (selected to span the range of quality classes) as training and validation data sets. We use 70\% of the set for training, and the remainder for validation. Figure~\ref{fig:perform} presents the performance (for both training and validation sets) of the three models of M1, M2 and M3, described in Figure~\ref{fig:combined}. For M1, the maximum accuracy of the validation set reaches $\sim92\%$; for M2 the peak validation is similar, while for M3 (the combined model) the peak validation is significantly higher at $97\%$. Considering the blue dashed line, M1, we see that there is a significant over-fitting effect after $\sim11$ epochs (the performance of the training set increases, however, there is no improvement for the validation set). The performance characteristics for M2 (the red lines) are different from M1, in which there is no significant over-fitting regarding this model. However, the performance is the same as M1 (after $\sim$13 epochs the accuracy is $\sim 92\%$). For M3, there continues to be improvement through to epoch 14 -- more than 97\% for epoch 14. After epoch 14, the performance of the network does not significantly change, and there is no strong evidence of over-fitting. (The validation and training sets track similar success rates.) M3 provides a superior quality classification -- the goal of our model.

The two sets of input data (i.e., Input-1 and Input-2, for M1 and M2) are informative and provide reasonable learning performance. As stated earlier, by itself the statistical information distributed on the SOM network has patterns that can be distinguished by a deep model. These patterns provided a classification accuracy of $\sim92\%$. However, a combination of the input information significantly increased performance. In this way, the information is mixed and enables more distinct patterns to be discovered, providing the remarkable improvement seen in M3 over M1 and M2. Besides, M3 provides smooth learning behaviour when compared to M1 and M2. We also see significantly better performance in epoch 1, which is a sign of better and more relevant information being provided to the deep model.  

It should be noted that the deep model (in M3) does not use any direct information from  SE such as ellipticity or ISO.  SOM presents suitable coordinates (RA and Dec in Fig. \ref{fig:combined})  to assemble the 20 small, cut-out images, along with associated statistical information, as the auxiliary data. Then we leave it to the CNN part to detect significant elongation or observation conditions, for example.

The stated performance of M3 is related to the validation set; to check this performance more robustly, we randomly selected different test sets and conducted a meticulous examination of the input images. These sets had not already been seen by the network (as training or validation sets), so they can provide a more robust proof than the 30\% validation set. In this respect, more than 1,500 exposures (i.e., more than 54,000 sub-images) were examined and then compared with the classification probabilities predicted by M3. Here we obtained even greater than 97\% accuracy, and we suspect that some of the images within our extensive training and validation sets may have been misclassified during the initial inspection, which involved the visual classification of over 100,000 images, due to fatigue during the classification effort. 

\subsection{Examples classifications}
\begin{figure}
\centering

\includegraphics[width=6.7cm,height=6.7cm,angle=0]{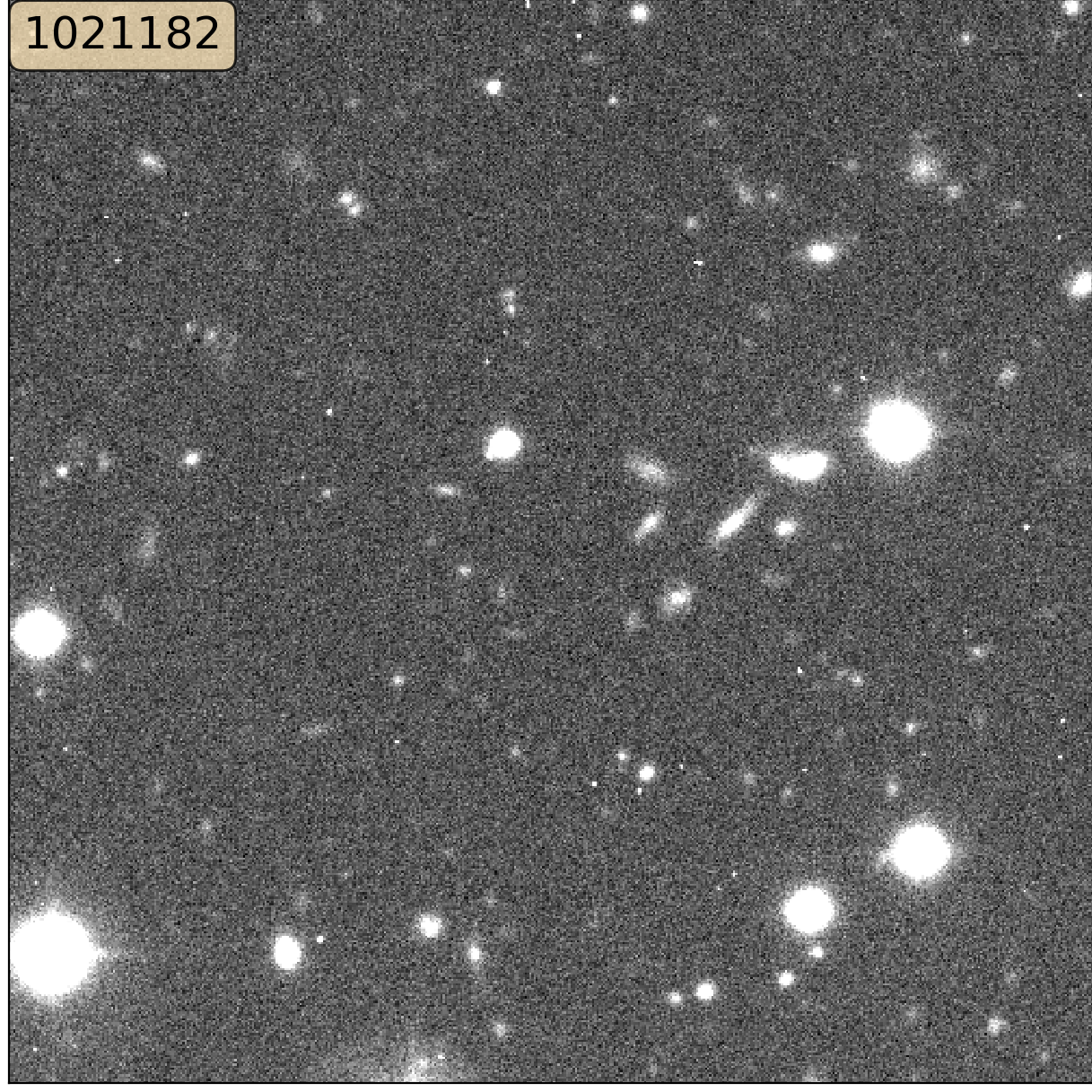}
\includegraphics[width=6.7cm,height=6.7cm,angle=0]{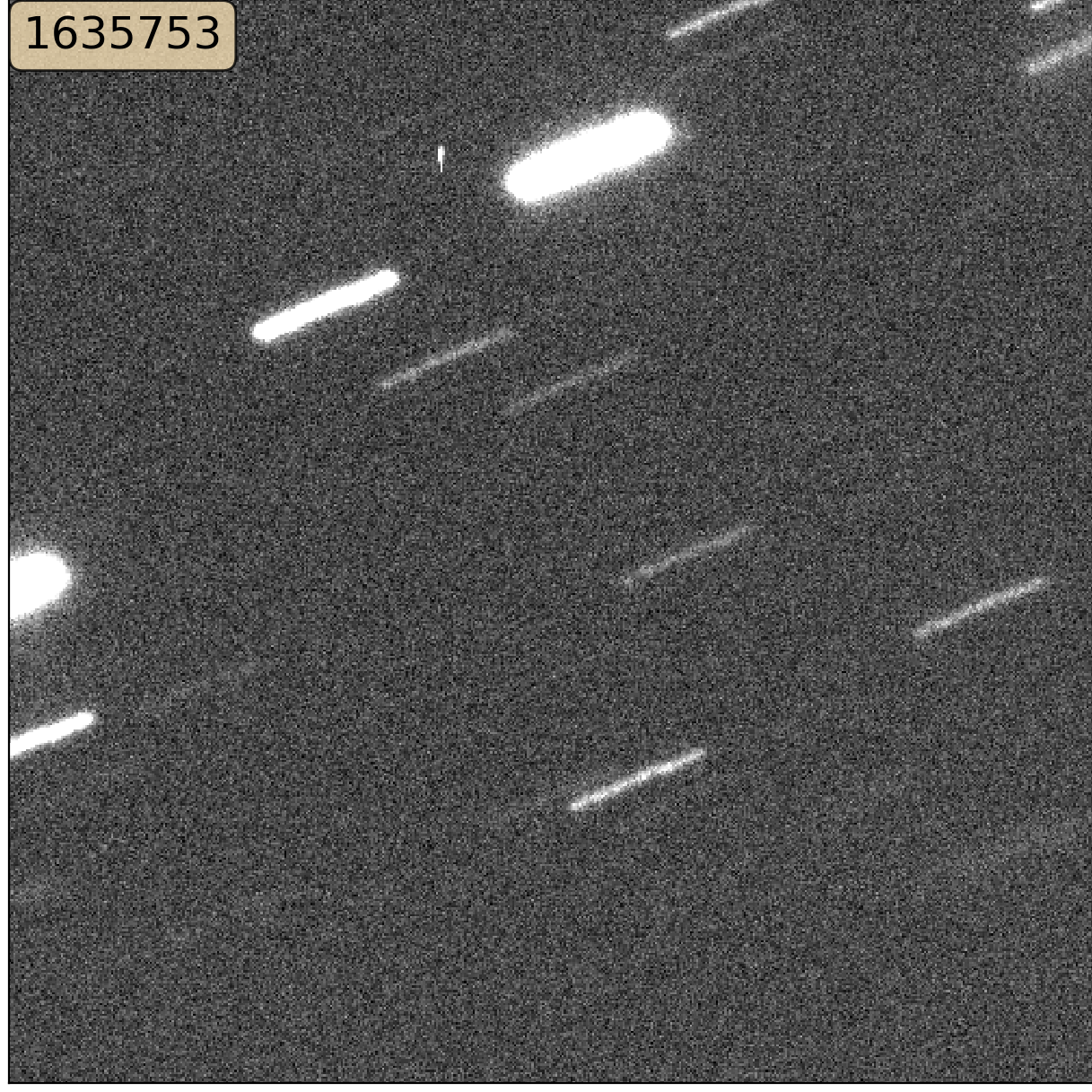}
\includegraphics[width=6.7cm,height=6.7cm,angle=0]{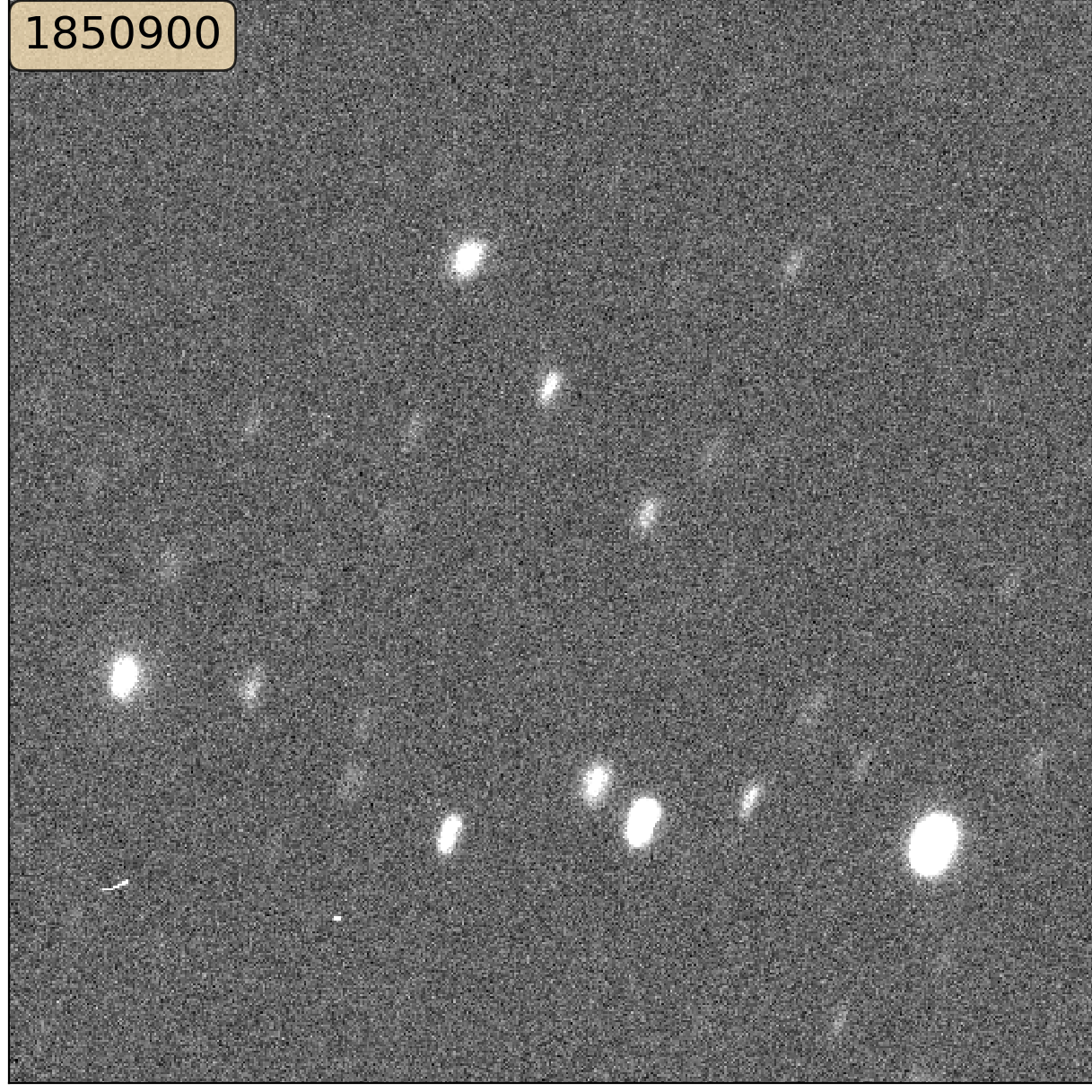}
\includegraphics[width=6.7cm,height=6.7cm,angle=0]{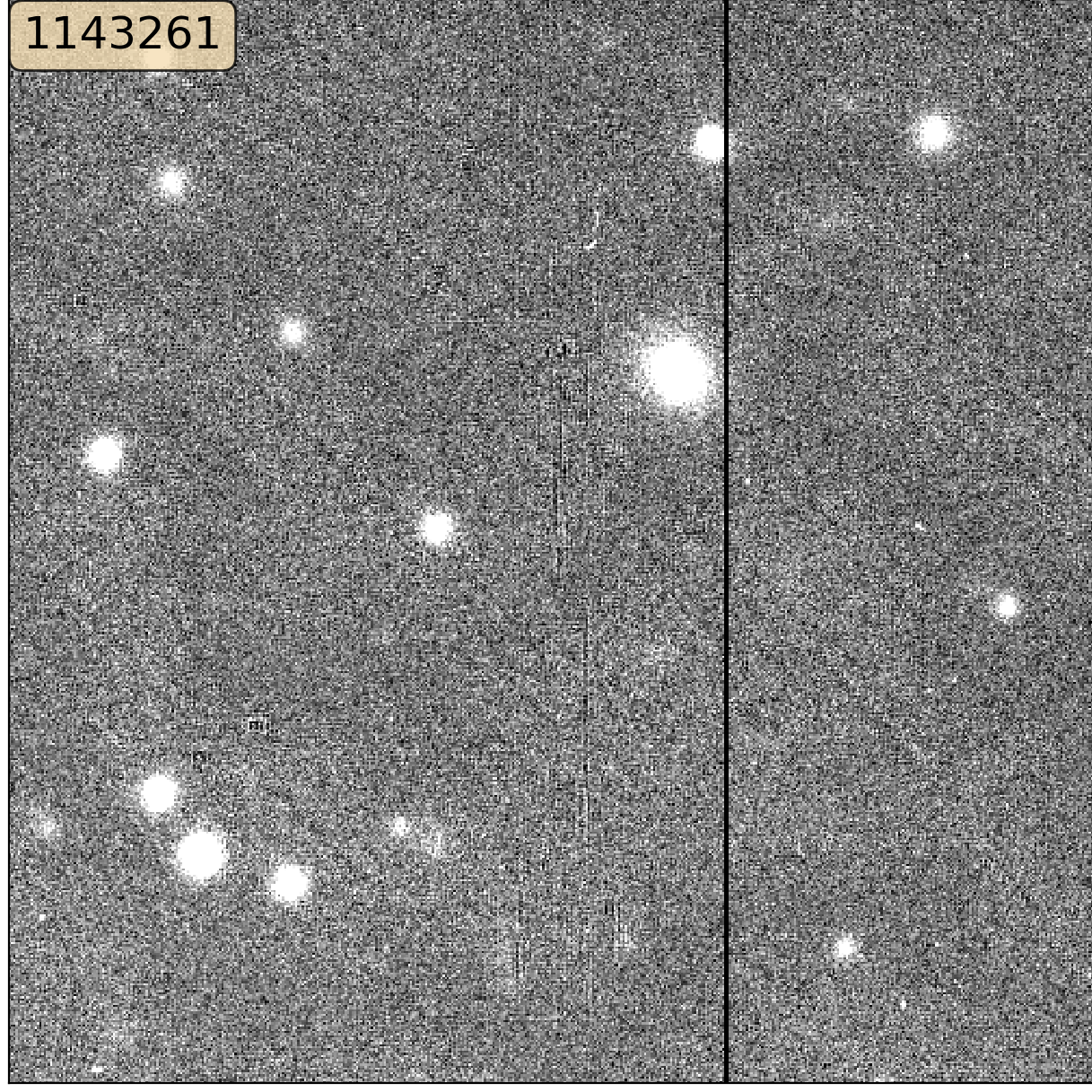}
\includegraphics[width=6.7cm,height=6.7cm,angle=0]{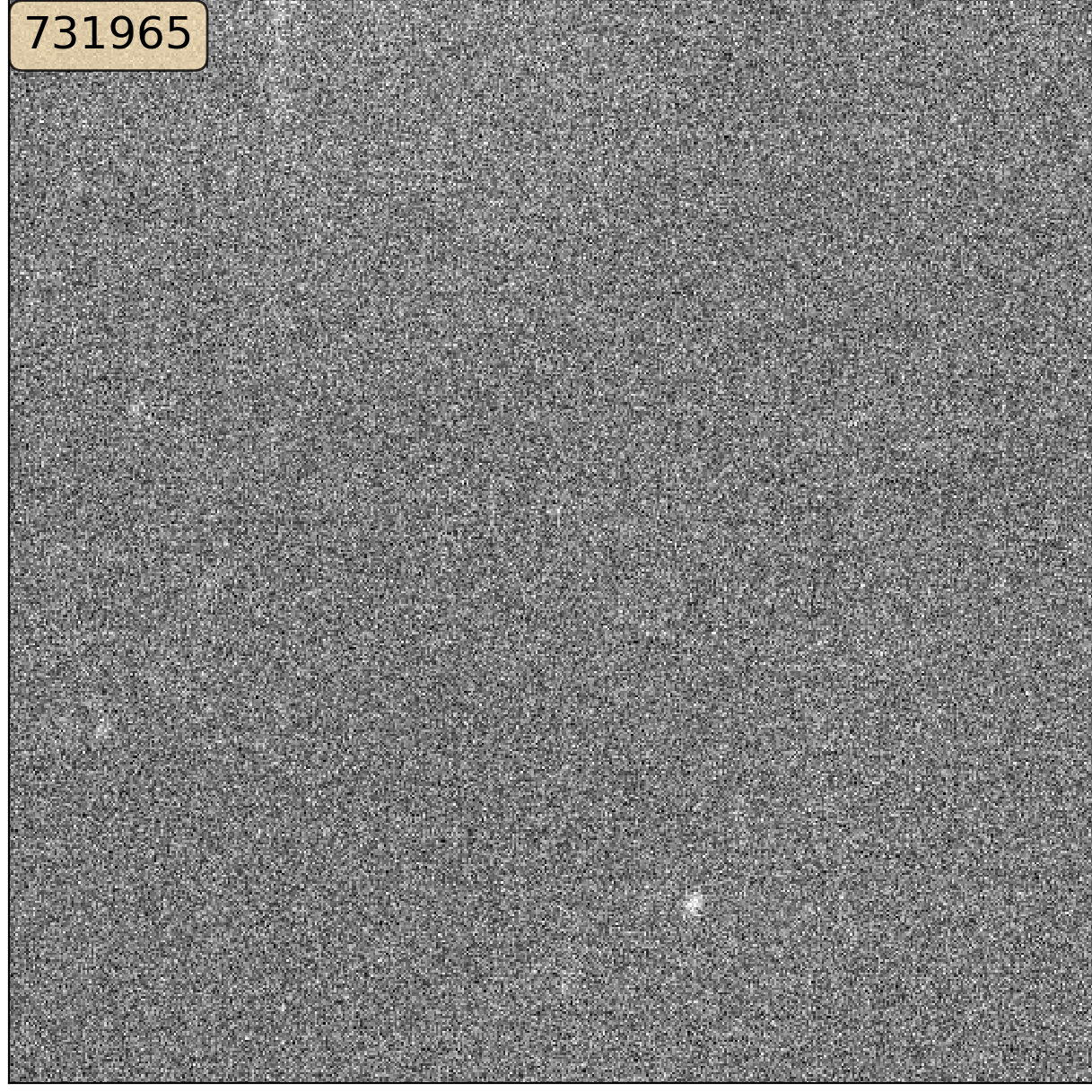}
\includegraphics[width=6.7cm,height=6.7cm,angle=0]{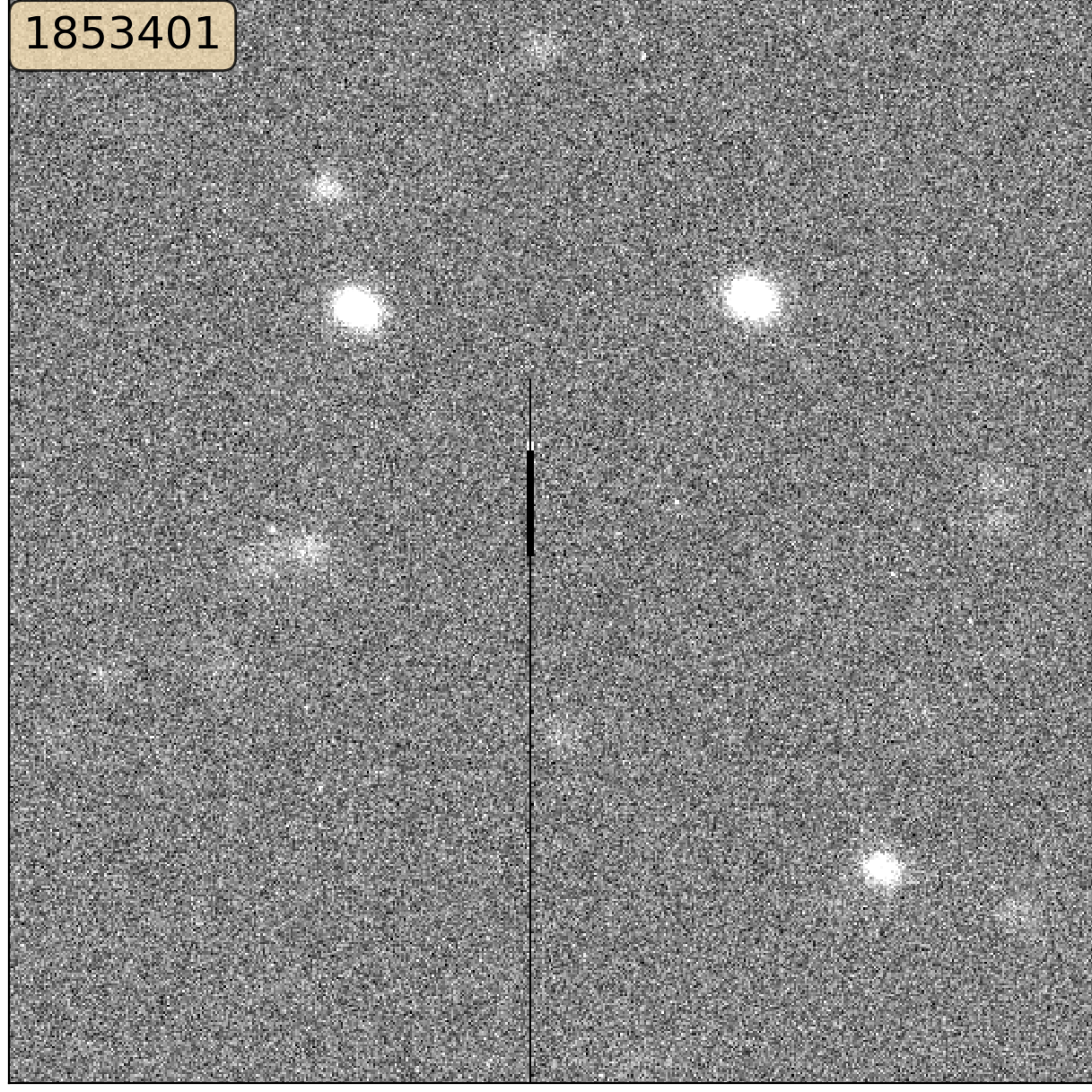}

\caption{The images that are listed in Table \ref{tab:t1}, with associated IDs: 1021182==Good, 1635753==RBT, 1850900==BT, 1143261== B-Seeing, 731965==BGP,  1853401== a combination of BT and B-Seeing. }
\label{fig:6tabel}
\end{figure}

Using deep model M3, we classified over 220,000 exposures (more than $\sim8,000,000$ images) in less than one day of computation. At no time did the performance of the process decay due to fatigue! Below we present some examples of these quality classified images. We present a table of some examples (exposures) with five different probabilities (and a column of the number of dead CDDs which might be found in an exposure) predicted by our method. For example, the first item of Table.\ref{tab:t1} (ID=1021182) shows that the exposure is a good one with more than 99\% probability \footnote{The images with different Observation IDs can be downloaded from http://www.cadc-ccda.hia-iha.nrc-cnrc.gc.ca/en/search/}. To illustrate the detail, only a part of this exposure is shown in the top left of Figure~\ref{fig:6tabel}. In the top right plot, we see an RBT exposure (1635753) with a probability more than 94\%. The RBT’s component of this exposure is powerful and completely over-shadows the other (Good, BT, B-Seeing, BGP) components. ID= 1850900 shows an exposure with BT, ID=1143261 is related to a B-Seeing image, and ID=731965  is an exposure with a problem in the background.  The last ID (1853401), in the table, show a mixed case. This exposure is a combination of BT and B-Seeing. 

\begin{table}
\centering
\caption{A sample of predicted probabilities by the method used in this paper}
\label{tab:t1}
\begin{tabular}{lcccccc}
\hline
ID & Good & RBT & BT & B-Seeing & BGP& $\rm{N_{Dead\_CCD}}$ \\
\hline
1021182  & 0.999 & 0.000 &     0.000 &     0.000 &     0.001 &   0\\ 
1635753  & 0.002 & 0.941 &  0.000 &     0.001 &  0.057 &  0 \\
1850900 &  0.141 &  0.016 &  0.843 &  0.001 & 0.000 &     0 \\
1143261 &  0.003 &  0.002 &  0.000 &     0.972 &  0.023 & 0\\
731965 &  0.000 &     0.002 &  0.000 &     0.001 &  0.997 &  1\\
1853401 &  0.065 & 0.000 &   0.260 &  0.673 & 0.002 & 0\\
\hline
\end{tabular}
\end{table}

\section{Discussion}
\label{sec:discussion}

To increase the density of useful pixel values being provided to the ML neural network, we select subsets of the image. Here, to make the processes easier for our ML algorithm, we provide subsets of specific forms of the image data (i.e., Input-1 and Input-2). The range or selection of these image subsets is determined using a SOM network with some selected input. Here we have iterated over a number of possible SOM network inputs and find that a SOM network with five source characteristics parameters allowed our deep model to distinguish between different image classes. Increasing the number of parameters does not improve the situation. Our process uses the source feature SOM to allow our model to be fed specific image subsets, ensuring that enough information is available to the model without overloading the input with repeated examples of similar information sets.

We found that for the MegaCam images being examined, a low exposure time value correlated strongly with poor image quality. We set all images with exposure times $>30$ seconds to 30 to group these longer exposures and allow the SOM to highlight the lower exposure times. Using exposure time is an example of using a metadata as a feature to a SOM. Using some expert knowledge derived from a careful examination of the input data can be critical to the success of the ML classifier.

Once the model was trained, determining image quality was quite rapid. The time required for preparation of the image subsets to be provided to the trained model is driven by the access and network speeds connected to the storage nodes that house the MegaCam images. A parallel computing system can increase the speed of data preparation. However, for our particular use case, the image assessment speed is driven by access to the stored image set. 
In other words, when an image and the relevant input catalog, from source extractor, becomes available on disk to the combined model the quality assessment requires less than 1 second of computing time. 
The number of clusters in a SOM is a tuneable option; here, we have classified the SOM into 20 clusters. We could use a 30-cluster SOM, for example, however, that would require increased data access to train and use the model. We found the 20-cluster SOM adequate for the classification and rapid enough to be practical given the computing and storage facilities at hand.

\section{The comparison with traditional measurements}
\label{sec:comparison}

\begin{figure}
\centering
\includegraphics[width=12cm,height=9cm,angle=0]{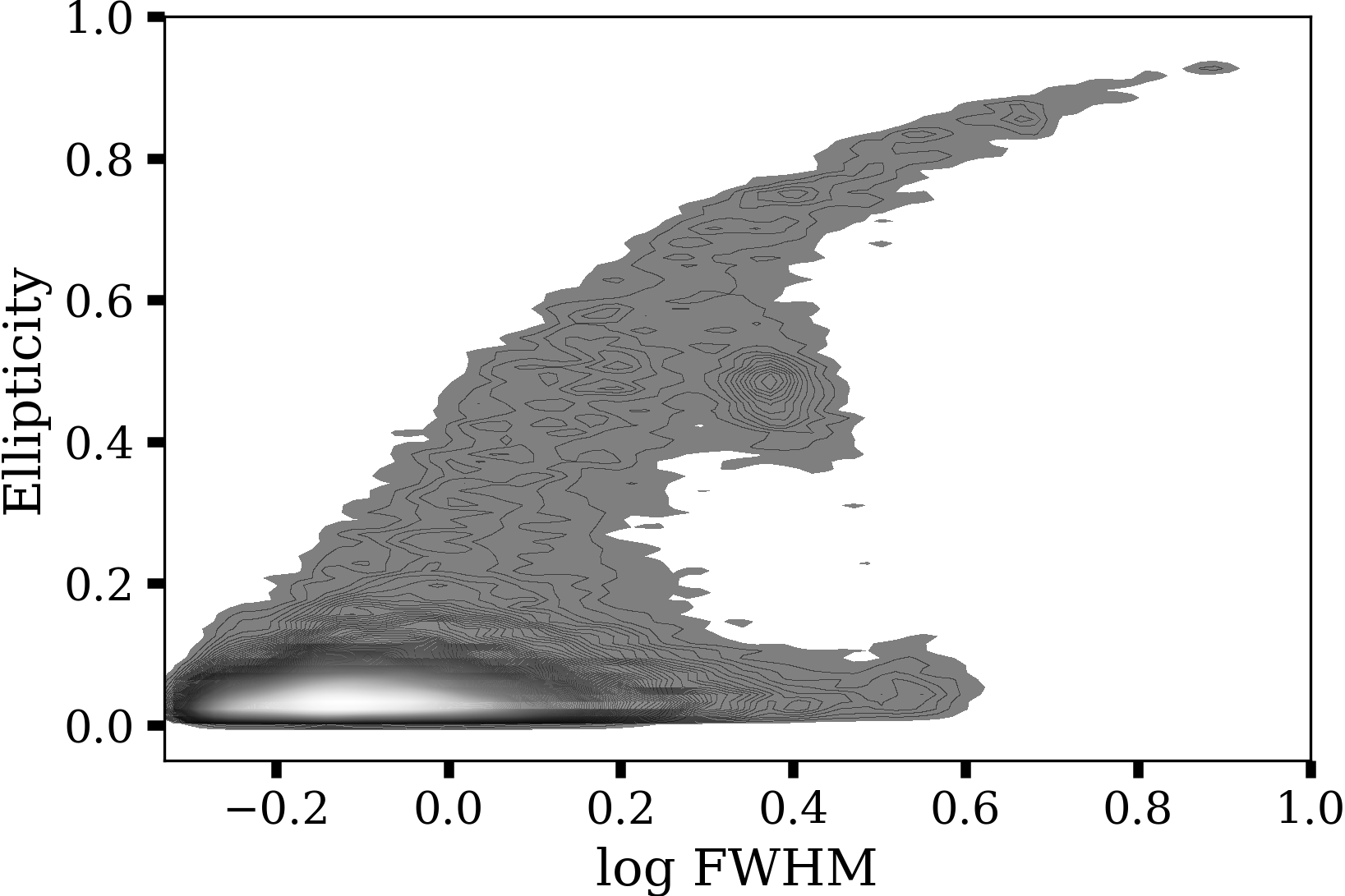}
\caption{Each point represents the average value of two parameters -- FWHM and ellipticity of available point rcerces -- in one image. The distribution of the average points for $\sim74000$ images is shown with a high-density region in the lower left of the plot.}
\label{fig-fwhm-ellip}
\end{figure}

As mentioned in Sec.\ref{introduction},  traditional methods, to assess the quality of images may use a variety of feature measurements, such as the mean PSF width and ellipticity of bright point sources of an image. To compare our result with conventional methods, we use 77,000 images (each with 36 CCDs), as a new test set. We use SE to measure the average FWHM and ellipticity for all bright point sources in an image. In Fig.\ref{fig-fwhm-ellip} we plot the distribution of the log of these average FWHM values against the average ellipticity for the 77,000 exposures. In the classical approach, one would select images with ellipticity $\lesssim 0.2$ and log(FWHM) $\lesssim 0.2$ as those with high-quality and we see from the figure that this is the region with the highest density of images in our ellipticity/FWHM space. This region can be considered as an area that selects out the highest quality images, i.e. those with no tracking issues and good observation conditions.  We show below that our ML approach finds images that are in this high-quality region but are, in fact, poor quality data.  First, we will present five plots to compare our algorithm with this classical method.

\begin{figure*}[ht]
\centering

\includegraphics[width=18.cm,height=9cm,angle=0]{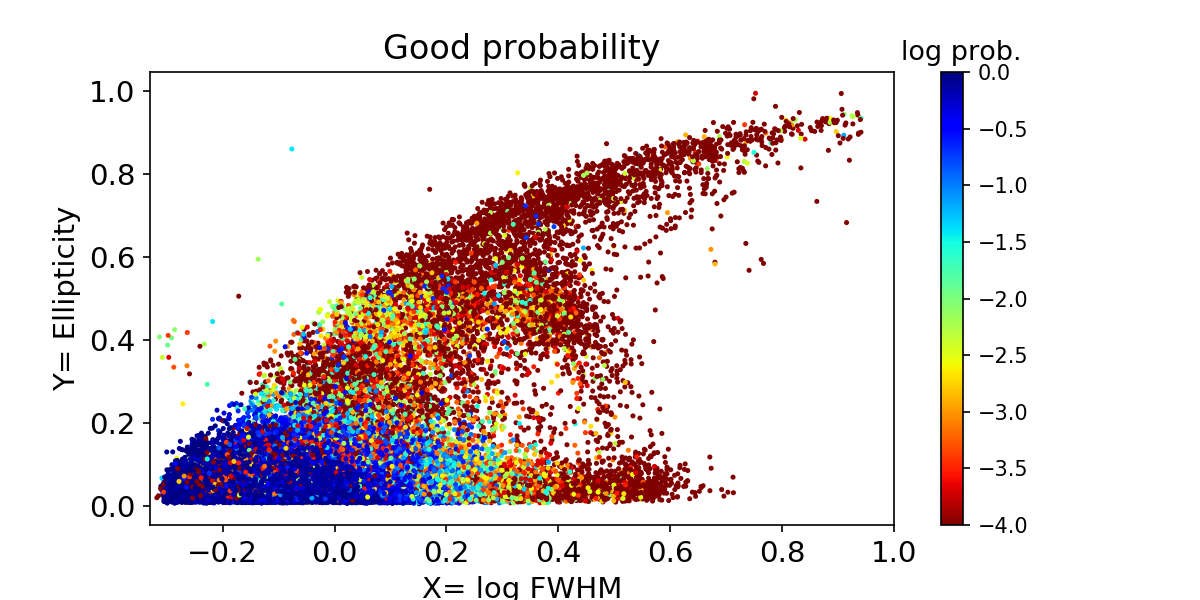}
\includegraphics[width=8.6cm,height=5cm,angle=0]{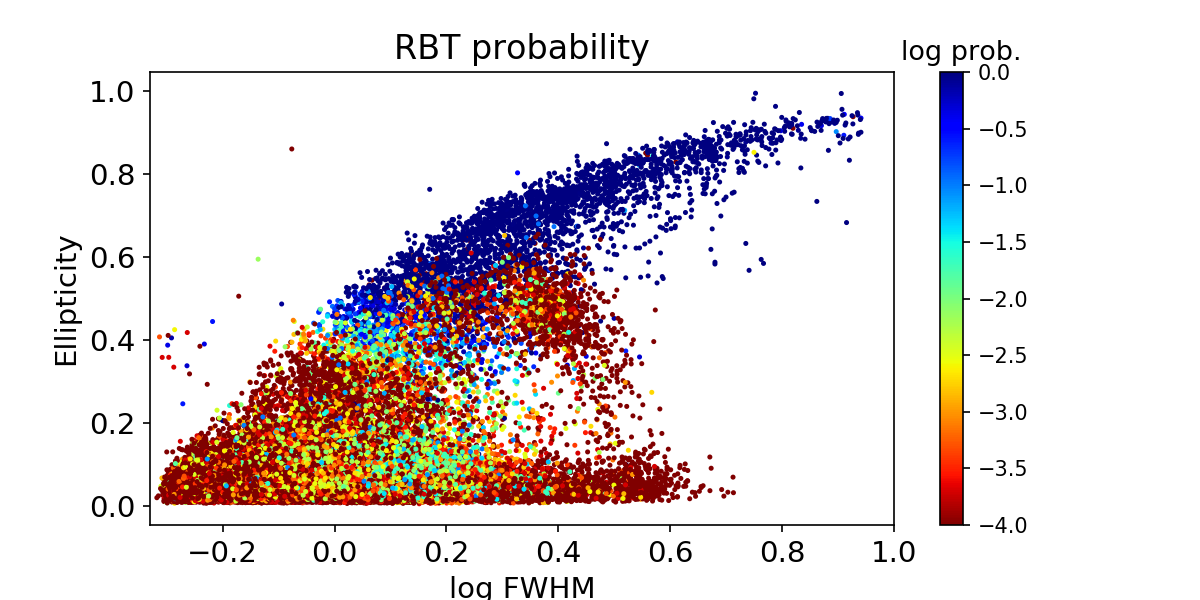}
\includegraphics[width=8.6cm,height=5cm,angle=0]{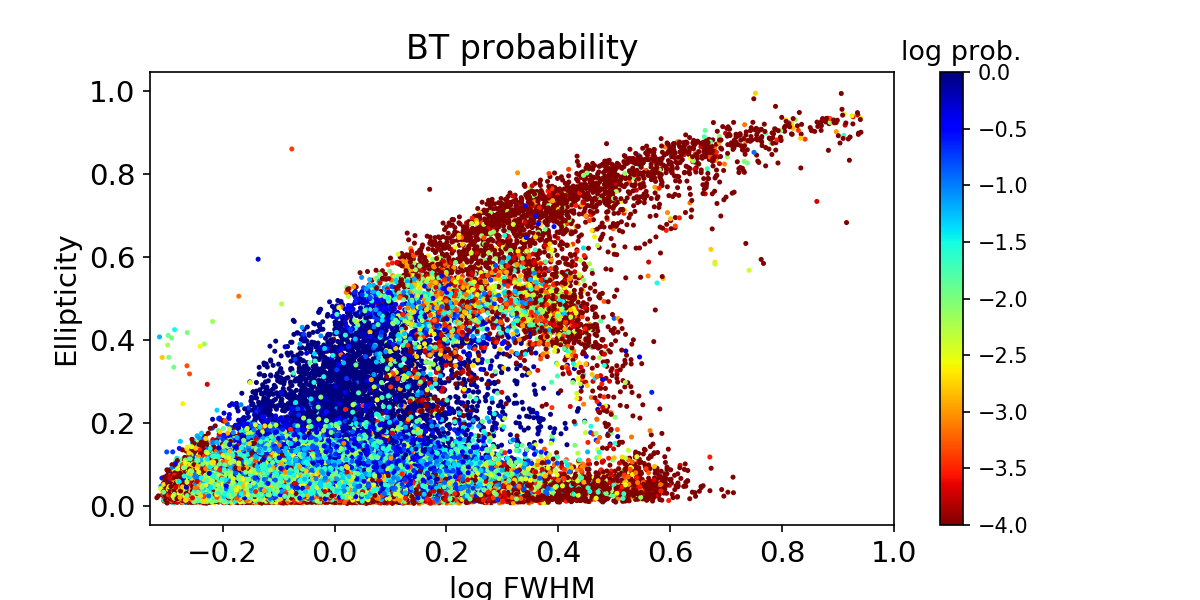}
\includegraphics[width=8.6cm,height=5cm,angle=0]{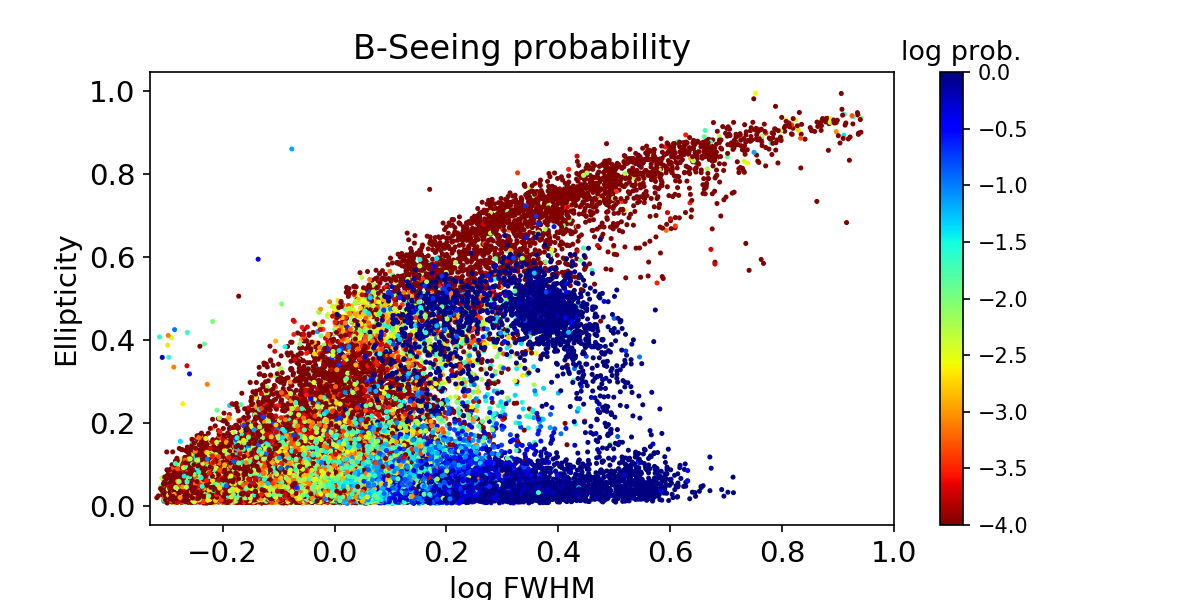}
\includegraphics[width=8.6cm,height=5cm,angle=0]{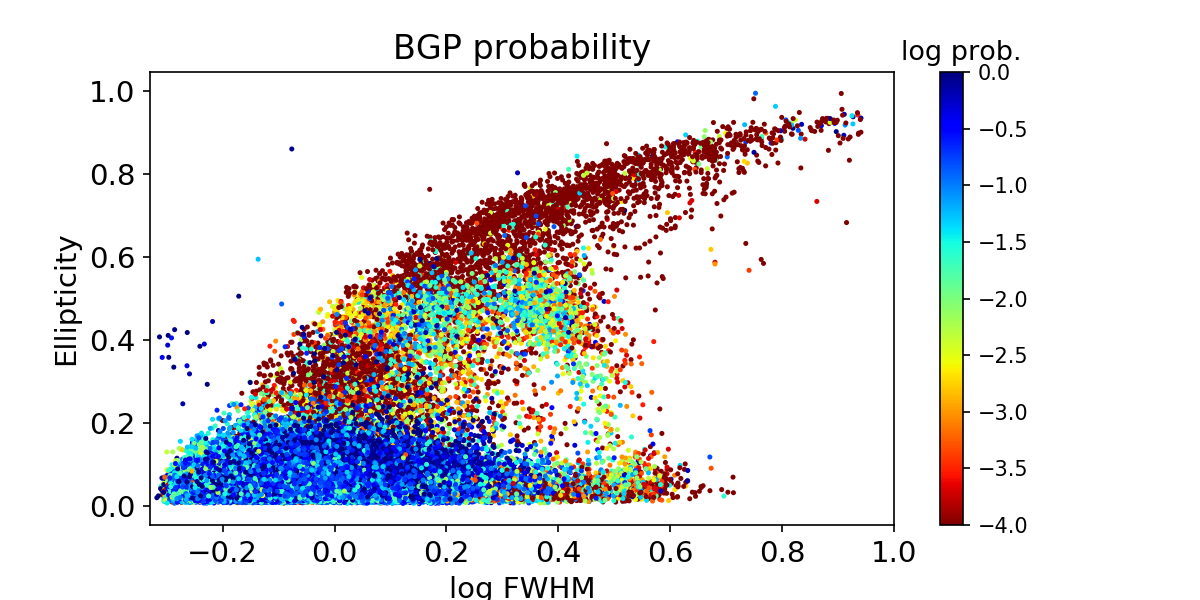}

\caption{The same as Fig. \ref{fig-fwhm-ellip}. The points are colour-coded by the five different probabilities obtained by our method. The top panel shows the probability based on the first, i.e., 'good' probability. The bluer points display the higher probabilities. The four other smaller plots show the other probability plots.}
\label{fig-5prob}
\end{figure*}

In Fig.\ref{fig-5prob} we show five different plots similar to Fig.\ref{fig-fwhm-ellip}, in which the points are colour-coded by the five different quality probabilities described in this paper. On the top panel, for example, we plot the value of the 'good' probability versus ellipticity and log(FWHM). As can be seen, images with high `good' probability are concentrated in the same dense region shown in Fig.\ref{fig-fwhm-ellip}. Generally, for binary classification, a natural decision boundary is 0.5 to separate two classes of a classification. Here we have a multi-class classification (five categories). Usually, the decision boundary for an image, for being good or bad, can be found after we get all the predictions (of the test sets) and after more visual inspections. We have found that a decision boundary of 0.20 for 'good' probability is a proper boundary with minimum confusion (i.e., log P$\gtrsim-0.5$, in the colour-coded bars). The area around the decision boundary is a little fuzzy and contains most of our misclassification cases. As can be seen in the top figure, there is a set of images for which the `good' probability is significantly below the good threshold but are still classified as high-quality images using the traditional approach.

As another example, the plot related to RBT, in Fig.\ref{fig-5prob}, shows that the images with significant bad tracking (high RBT probability, blue) are nearly absent from the high-quality region of the classical method. The high-probability RBT cases are related to the area that has high ellipticity and FWHM, but some low-probability RBT cases are also in this zone.  These low-probability RBT measures turn out to select for images with poor focus. Thus, the RBT classification criterion provides additional diagnostics about the cause of the poor quality data.

The BT  parameter has a high-probability for images where the tracking is poor, but not catastrophic as is the case for high-probability RBT images.  Looking at Fig.~\ref{fig-5prob} one can see that the 'ridge' of images in the high-ellipticity large-FWHM section of the plot images transition from high-RBT probability (very large FWHM) to high-BT probability as moderately large FWHM.  There is a continuous connection between the RBT and BT probabilities.  Using both parameters allows selecting between catastrophic and poor image cases.  

The plot B-Seeing has two populated areas. One area is related to high FWHM with low ellipticity are images that are poor due to poor seeing while the other is associated with high FWHM and high ellipticity and selects out those images where the focus was poor.  Filtering out images with a high probability of B-Seeing removes both of these cases.

\begin{figure*}
\centering
\includegraphics[width=6.7cm,height=6.7cm,angle=0]{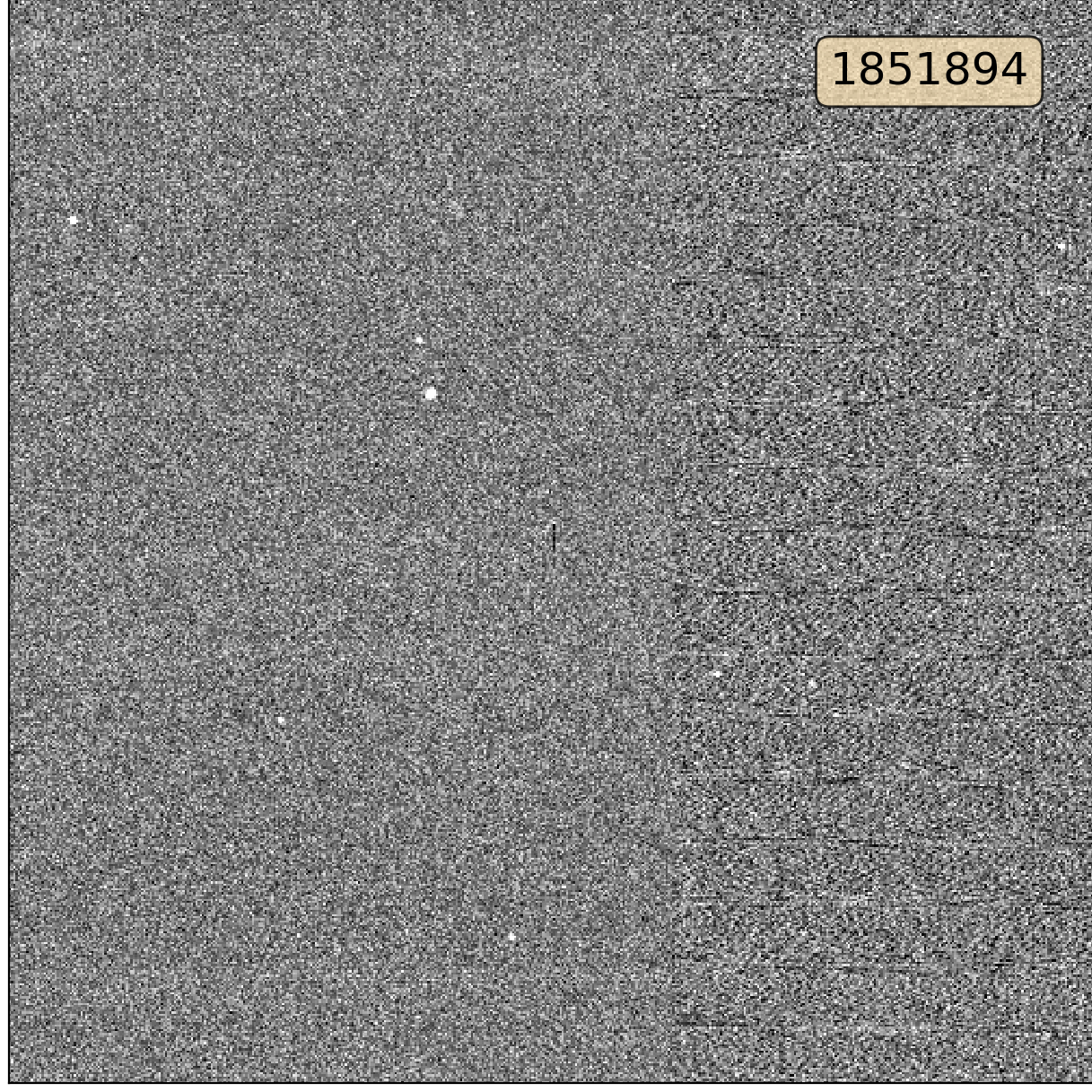}
\includegraphics[width=6.7cm,height=6.7cm,angle=0]{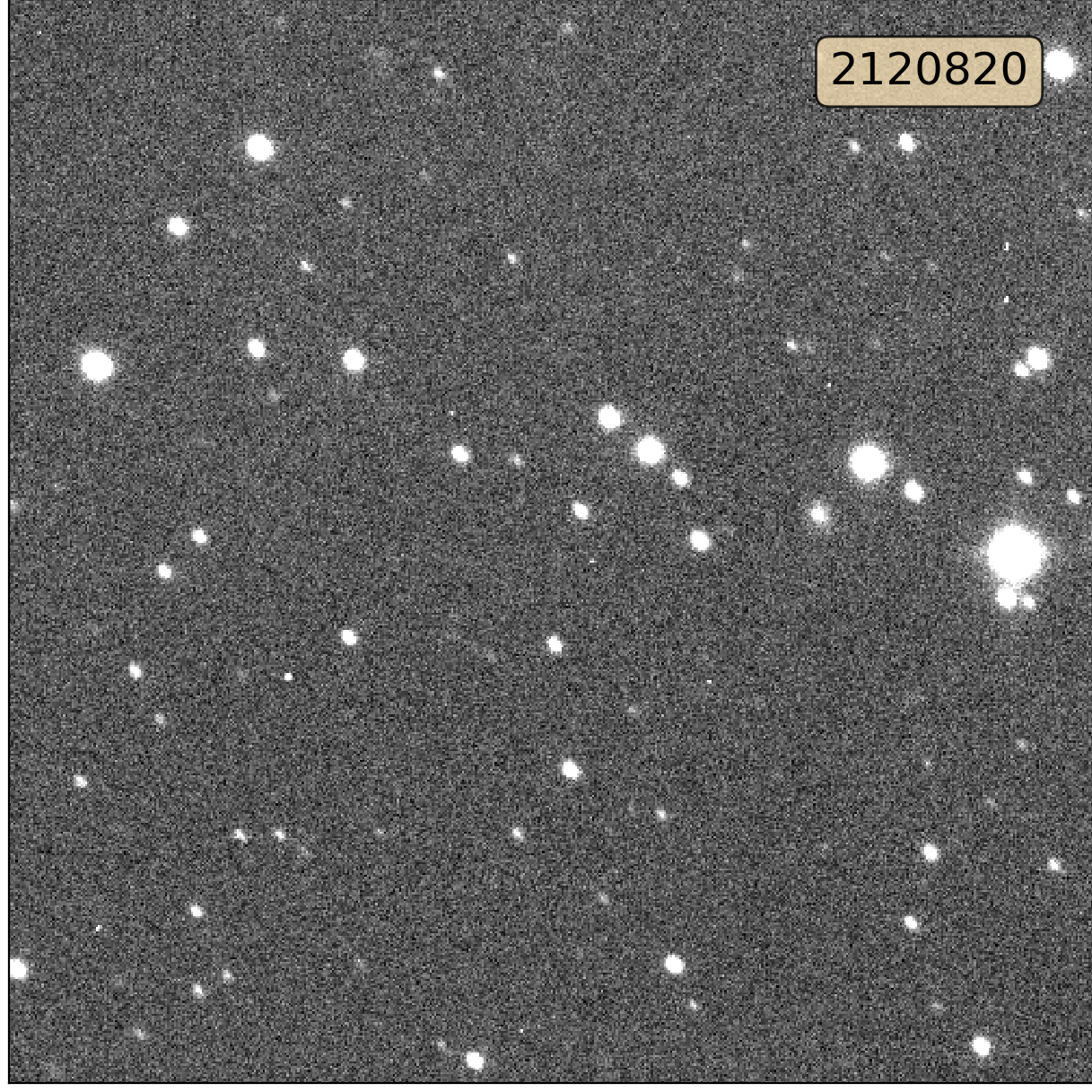}
\includegraphics[width=6.7cm,height=6.7cm,angle=0]{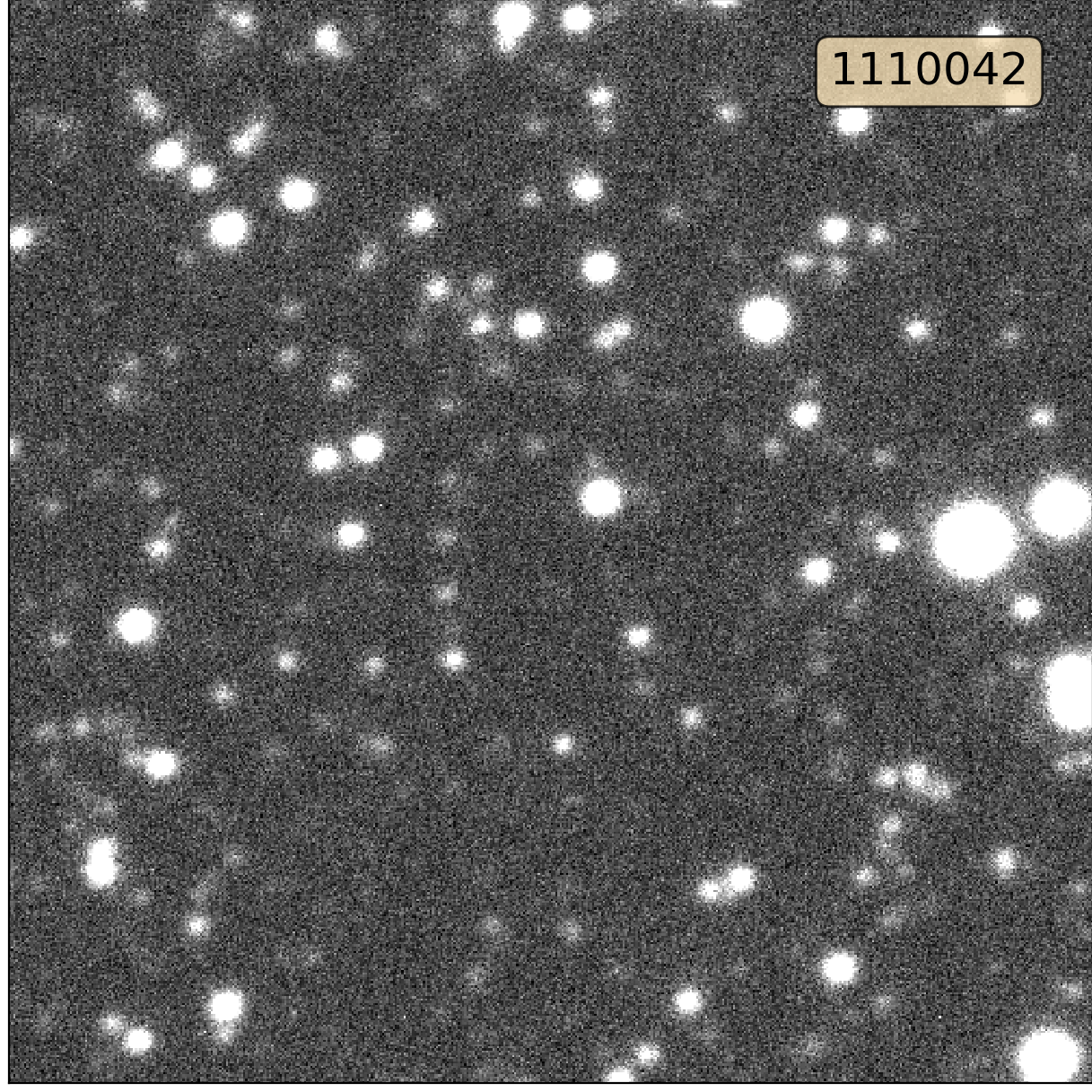}
\includegraphics[width=6.7cm,height=6.7cm,angle=0]{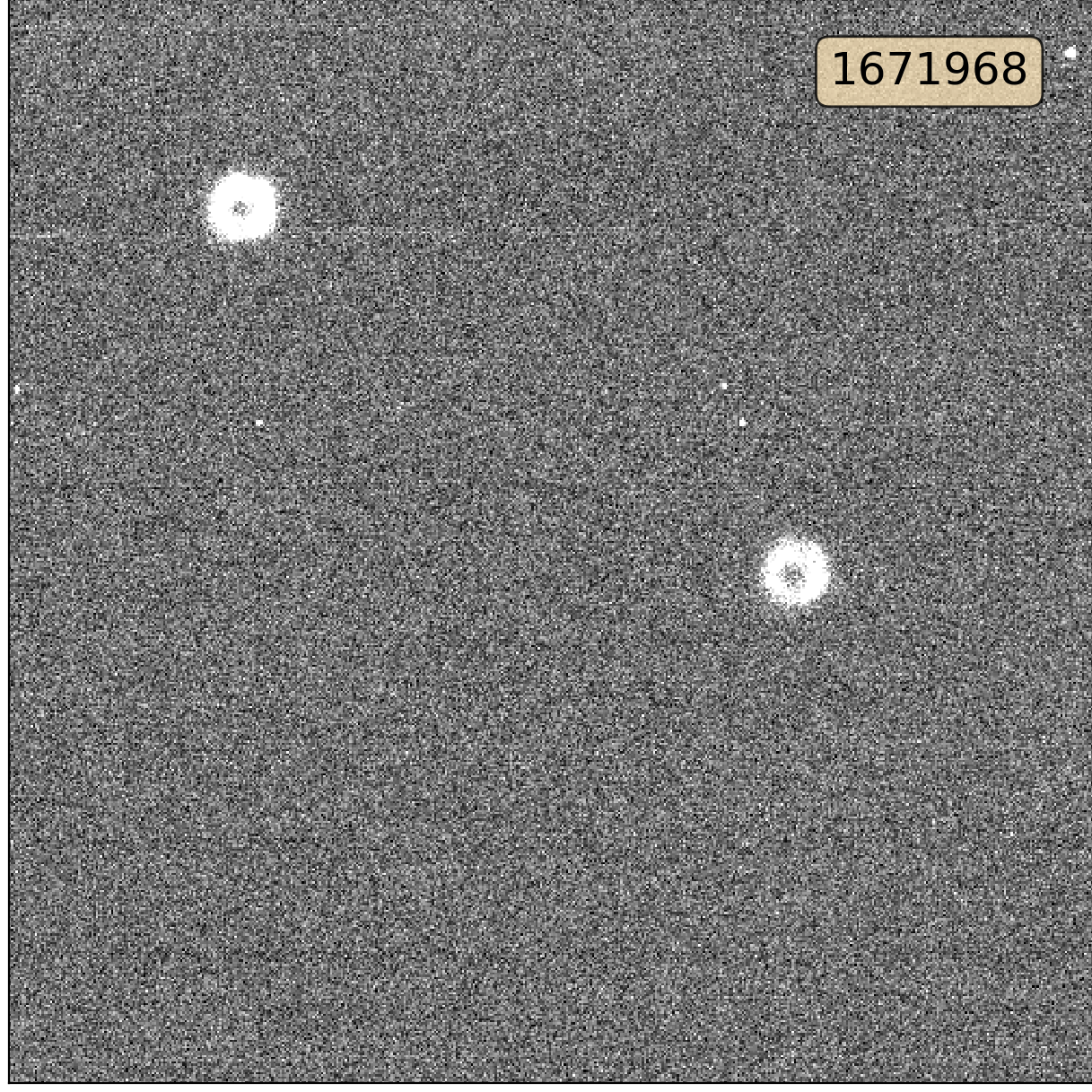}
\includegraphics[width=6.7cm,height=6.7cm,angle=0]{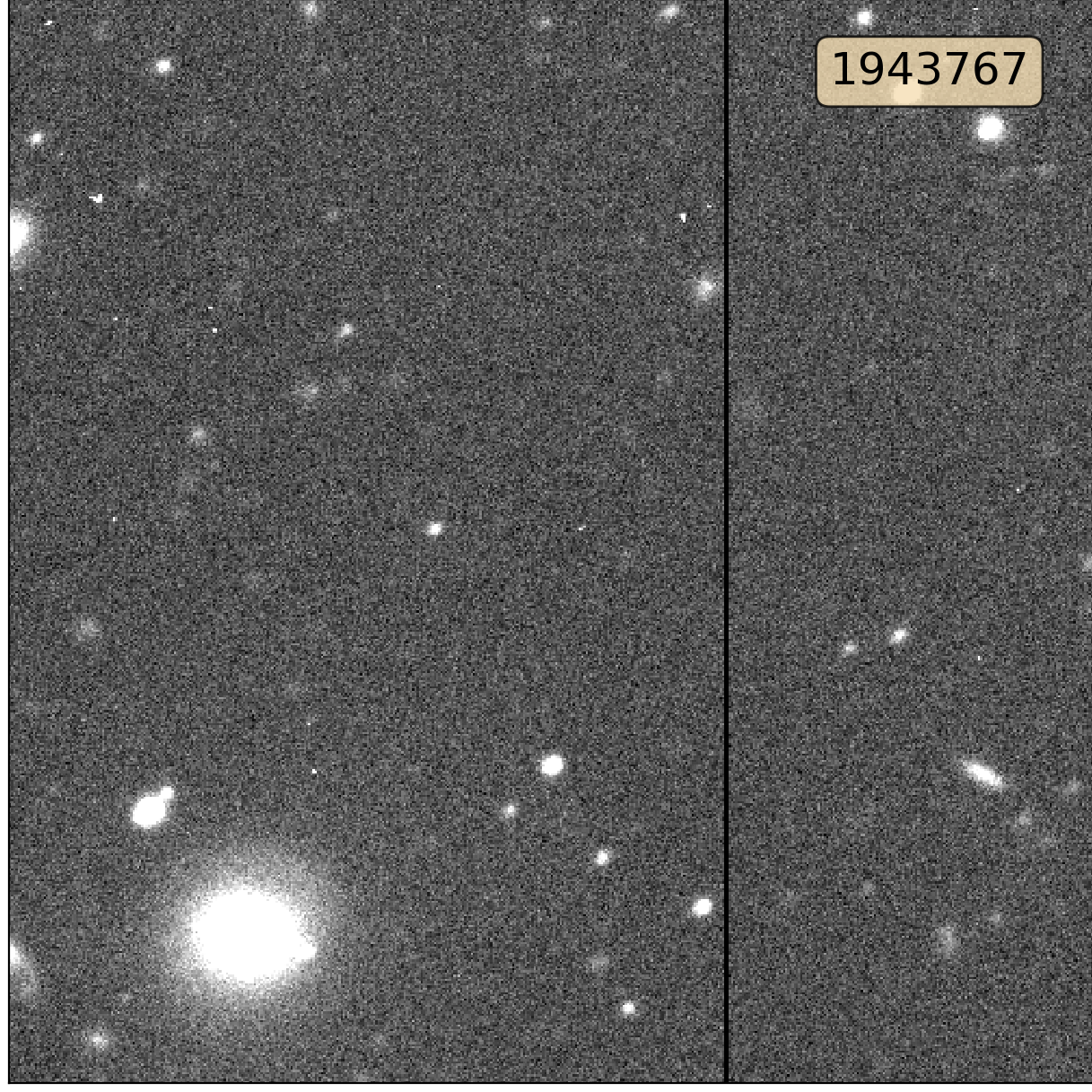}
\includegraphics[width=6.7cm,height=6.7cm,angle=0]{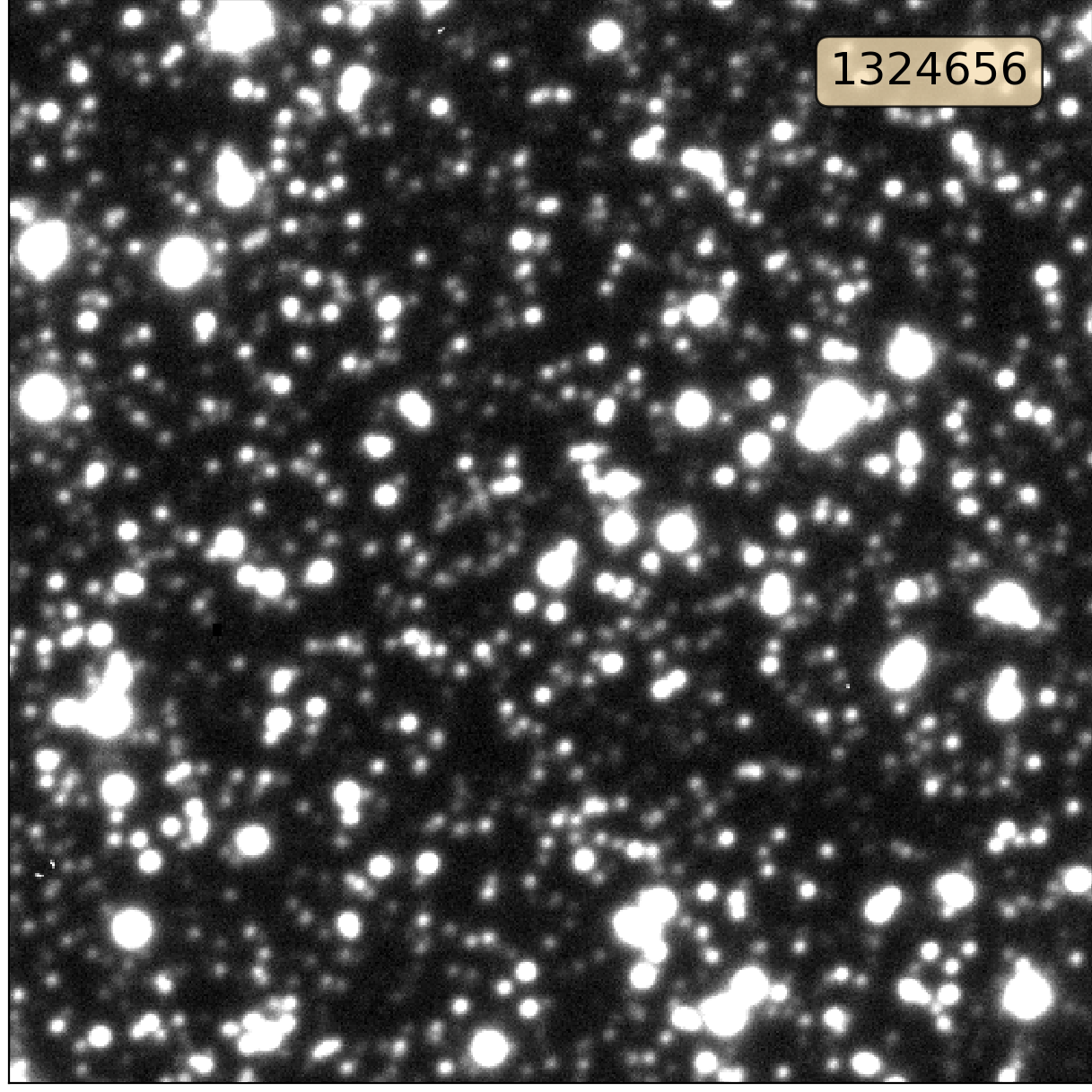}
\caption{In the plot, we show six different examples of a test set with  77,000 exposures (each with 36 CCDs). The IDs denoted in the images with associated probabilities are described in Sec. \ref{sec:comparison}. The images with IDs 1851894, 2120820 and 1110042 have acceptable ellipticity and FWHM; however, our models detect some problems with images such as BGP, and the combination of B-Seeing and BT. The Image with ID =1671968 is an example of an image with an out-of-focus character which is detected by the method as B-Seeing. The two lower images have been classified as high-quality images; however, with small detected problems such as slight bad tracking for ID=1943767 and a combination of small probabilities of BT, B-Seeing and BGP for  ID= 1324656}.
\label{fig-newtest}
\end{figure*}

Finally, the plot of BGP probability shows that some images in the high-quality section of FWHM/Ellipticity space have high BGP probability, i.e., P(BGP $>0.8$ (very dark points). These images exhibit problems with the sky background, for which ellipticity and FWHM are not diagnostic (see the top-left panel of Fig.~\ref{fig-newtest} for an image with P(BGP) $>0.8$ but FWHM/ellipticity in the high-quality zone).  

The reader is cautioned against examining these probability plots independent of each other.  Each case is treated as exclusive; the sum of the probability of the five characteristics is 1.  Thus, images for which one characteristic (such as BGP for example) is nearly 1, the other probabilities are necessarily near 0. For example, we may have a problem in the background for images with high RBT probability as the prominent component, and thus the BGP probability will be low (sources with high ellipticity and low BGP in the BGP plot are generally representative of this situation).  A strong effect in one characteristic dilutes the impact of the others. Dominantly, for images taken under condition of RBT, the weather condition (which might create B-Seeing or BGP) is not significant to the quality assessment.

Fig.\ref{fig-5prob} reveals some low probability (i.e., red) points in high-quality part of the ellipticity-FWHM space. That means some images do not have a good predicted quality using these classical parameters.  Of the 70,000 images examined in this test, 7,400  images ($\sim10\%$) can be found with low probabilities of being good images (i.e., $P_{\rm{Good}}<0.2$) but with ellipticity $<$ 0.2 and FWHM $<$ 1.5 (classically considered good images).  In other terms, the traditional ellipticity/FWHM approach misclassifies 10\% of images.

Of the 7,400 incorrectly classified images, about 4,500 are associated with different problems in the background, based on the BGP probability, such as ID=1851894 with BGP$>0.92$ in Fig.\ref{fig-newtest} and also ID=731965 in Fig.~\ref{fig:6tabel}.  Visual inspections of a random selection of these 4,500 images reveal, universally, problems in the sky background and unusual sky patterns.  These problems were not detectable using traditional methods.  These background issues were not necessarily local fluctuations in the sky background but often low-level patterns in the image background. As another example, there are more than 650 of the 7,400 images that exhibit very minor issues with tracking (such as ID = 2120820, BT$>0.80$). More than 120 images in the high-quality zone but classified as poor images have a combination of bad seeing and bad tracking (such as ID=1110042 with probabilities Good= 0.1183, RBT=0.0790, BT=0.2514, B-Seeing= 0.3970, BGP=0.1542) which combine to remove them from the good image list. The examples show that the combined models can detect different issues that might be hidden for traditional methods. In the following, we show other interesting circumstances.

There are also many images of high quality that show minor traces of various 'bad' characteristics. For example, there are more than 3,000 images with $P_{\rm{Good}}>0.2$ which show some bad tracking probability (such as  ID= 1943767 with probabilities Good=0.3438, RBT=0.0213, BT=0.4233, B-Seeing=0.0158 BGP=0.1958) but the `Good' characteristic is the dominant one, and these images are selected as good. Also, there are different examples of high-quality images with a combination of little bad seeing, bad tracking, and BGP (such as ID=1324656 with probabilities Good=0.2360, RBT=0.0932, BT=0.2358, B-Seeing=0.2239, BGP=0.2111). Visual inspection of these images reveals that they are, indeed, suitable for use in our image stacking system. The mentioned examples can be easily classified from databases for further investigations.  In fact, one powerful aspect of this work is that the classifier can find combined problems and present them to users in terms of different probabilities. In other words, one can use a combination of probabilities to search for the desired image in terms of quality. 

Many unique images are detected as bad seeing images; however, when they are inspected, they show that the images have out-of-focus characteristics. These images are mostly located in the high FWHM and high ellipticity region indicated in the plot B-Seeing (such as ID= 1671968, B-Seeing $\sim1$). The ability to distinguish between poor-focus images and images taken in poor-seeing could be used as a feed-back into telescope operations if the model were deployed at the observatory.

Finally, there are also many images of very crowded stellar fields (with huge deblending influences) which may be flagged as BGP.  These images are quite rare and indicate that the model can detect problems that may not have been considered in our training set. 

\section{Summary}
We present a method in which two groups of input data are fed to a deep neural network to classify complex, ground-based telescopic images. As an example of a complex data set, we use CFHT MegaCam images to explore and demonstrate our approach. The first input contains the pixel information of the images, which we call representative images. They comprise a small set of cut-out sources obtained from a list that, in turn, is provided by the Self-Organizing Map method. This method allows us to cluster detected sources in an image and pick suitable representative sources. We show that for astronomical images we do not need to provide the entire (often large) image to a deep model, nor do we reduce the resolution of the image, which would remove useful information.   

We have tested using different, independent, sets of exposures and have found that a decision boundary of Good $>$ 0.20 provides an accuracy of more than 97\%, where the 97\% of images our algorithm classifies as ‘good’ are also classified as such during a visual inspection.  We suspect that cases of disagreement between the algorithm and the visual inspection step are driven by fatigue during the visual inspection step.  In addition, about 10\% of images classified as good using traditional ellipticity/FWHM criterion are, in-fact, not good images when classified with using the ML approach described here. 

\label{sec:conclusion}

\section*{Acknowledgements}

H.T. thanks Patrick Dowler for his support and help in using different services in CADC.

\bibliography{main}{}
\bibliographystyle{aasjournal}



\end{document}